\documentclass[12pt]{article}
\usepackage[utf8]{inputenc}
\usepackage{amsthm, amsfonts, amssymb, amsmath,enumitem, bbm, mathabx}
\usepackage{mathtools,natbib}
\usepackage{ragged2e,setspace}
\usepackage{graphicx}
\usepackage{hyperref}
\usepackage{geometry, xcolor}
\usepackage{fancyhdr}
\usepackage{multirow, booktabs}
\usepackage[ruled,vlined]{algorithm2e}
\usepackage{xr}

\def\boe{\begin{enumerate}}
\def\eoe{\end{enumerate}}

\newtheorem{theorem}{Theorem}[section]
\newtheorem{lemma}[theorem]{Lemma}
\newtheorem{proposition}[theorem]{Proposition}

\newtheorem{definition}[theorem]{Definition}

\newcommand{\E}{\mathbb{E}}

\newtheorem{assumptions}{{\bf Assumption}}

\newcommand\ca[1]{{\cal{#1}}}
\newcommand\lo[1]{_{{#1}}}
\newcommand\hi[1]{^{{#1}}}

\def\cid{\stackrel{\mbox{\tiny $\cal D$}}\longrightarrow}

\def\cip{\stackrel{\mbox{\tiny $P$}}\rightarrow}

\def\E{\mathbb{E}}

\def\L{{\cal L}}

\def\inv{^{\mbox{\tiny $-1$}}}

\newcommand{\ip}[2]{\left\langle #1,#2\right\rangle}

\def\sign{\mathrm{sign}}

\def\cid{\stackrel{\mbox{\tiny $\cal D$}}\rightarrow}

\def\ali{&\,}

\def\real{{\mathbb R}}

\def\ka{\kappa}

\def\L2T{L \lo 2 (T)}
\def\L2TX{L \lo 2 (T\lo X)}
\def\L2TX{L \lo 2 (T\lo Y)}

\def\ali{&\,}

\def\ali{&\,}

\def\eod{\end{document}}

\newfont{\rsfsten}{rsfs10 scaled 1100}
\newfont{\rsfstena}{rsfs10 scaled 800}
\newfont{\rsfstenb}{rsfs10 scaled 500}

\def\E{\mathbb{E}}

\newcommand{\ben}{\begin{enumerate}}
	\newcommand{\een}{\end{enumerate}}
 \newcommand{\PP}{\mathbb P}

\newcommand{\Hc}{\mathcal H}

\begin{document}

\title{\Large A Test for Treatment Heterogeneity under a Distributional Difference-in-Difference Framework\thanks{Lingzhou Xue is the corresponding author (Email: lzxue@psu.edu).}
}

\author{\normalsize
Satarupa Bhattacharjee$^\dagger$, Bing Li$^\ddagger$, and Lingzhou Xue$^{\ddagger}$  \\ \normalsize
			$^\dagger$Department of Statistics, University of Florida\\ \normalsize $^\ddagger$Department of Statistics, The Pennsylvania State University
            }

\date{}

\maketitle

\begin{abstract}
We develop a novel distributional Difference-in-Differences (DiD) framework to capture treatment heterogeneity across outcome distributions. By leveraging optimal transport, we use the control group to estimate the untreated distributional drift from the pre- to post-treatment period and apply it to the treated group's pre-treatment baseline, constructing a counterfactual distribution under the assumption of no treatment effect. We frame the null hypothesis as a distributional equality between the transported counterfactual distribution and the observed treated post-treatment distribution, and test it using a maximum mean discrepancy statistic in a reproducing kernel Hilbert space (RKHS). The resulting nonparametric omnibus test is sensitive to changes in location, scale, shape, and tail behavior. Under the null, we derive the asymptotic Gaussian quadratic-form limit of the test statistic, while under local alternatives, we provide a unified characterization of power that establishes its Pitman local power and moderate-deviation consistency. Our theory reveals how detectability is shaped by the interaction between transport-induced drift and RKHS geometry. Simulations and an application to the Card--Krueger minimum-wage data demonstrate that the proposed method identifies key distributional treatment effects missed by classical mean-based DiD.
\end{abstract}


\setstretch{1.3}
\section{Introduction}
\label{sec:intro}
Difference-in-Differences (DiD) is a cornerstone design for causal inference with repeated outcomes, widely adopted due to its conceptual simplicity and robustness to time-invariant unobserved heterogeneity. In the canonical two-period, two-group setting, let \(D\in\{0,1\}\) denote treatment group assignment, with the treated group \(D=1\) and the control group \(D=0\), and let \(t\in\{0,1\}\) index the pre- and post-treatment periods. For each unit, let \(Y_t(0)\) and \(Y_t(1)\) represent the potential outcomes under control and treatment at time \(t\). The observed outcomes are realized as $Y_0=Y_0(0)$ and $Y_1=Y_1(D)$. The primary causal estimand of interest is the average treatment effect on the treated (ATT), defined as
\(
\tau_{\mathrm{ATT}}
=
\E\!\left[Y_1(1)-Y_1(0)\mid D=1\right].
\)
Under the parallel trends assumption that
$\E\!\left[Y_1(0)-Y_0(0)\mid D=1\right]
=
\E\!\left[Y_1(0)-Y_0(0)\mid D=0\right]$,
the ATT can be identified through the familiar difference-in-differences representation:
\begin{equation}\label{eq:att}
\tau_{\mathrm{ATT}}
=
\bigl(\E[Y_1\mid D=1]-\E[Y_0\mid D=1]\bigr)
-
\bigl(\E[Y_1\mid D=0]-\E[Y_0\mid D=0]\bigr).
\end{equation}
The classical two-way fixed-effects model
\(
Y_{it}=\alpha_i+\lambda_t+\tau D_{it}+\varepsilon_{it}
\)
encodes this identification strategy algebraically: untreated potential outcomes evolve through additive unit and time effects, and any excess post-treatment change for the treated group is attributed to treatment.

This classical formulation is focused on mean effects, assuming the target of interest is a location shift or a contrast in conditional expectations. In many applications, however, the treatment changes more than the mean, as it may alter the spread, skewness, tail behavior, modality, or underlying dependence structure of the outcome distribution. {This setting aligns with the growing literature on
distributional data analysis, where probability distributions are treated as statistical objects \citep{petersen2019frechet,dubey2020functional,chenLinMuller2023wassersteinRegression,
zhu2023autoregressive_ot,zhang2024dimensiona,zhang2024nonlinearb,bhattacharjee2025nonlinear,iao2025deep,fontaine2026bayesian}.} Here, mean contrasts fail to capture substantive heterogeneity. For example, a treatment may leave the mean unchanged while reallocating mass across the support, or two interventions may yield identical averages with vastly different distributional shapes. Such insights have driven the DiD literature to transition from additive mean-shift models to distributional frameworks. 

Moving beyond mean-level contrasts, the Changes-in-Changes (CiC) model \citep{AtheyImbens2006} identifies the counterfactual distribution of untreated outcomes for the treated group. By assuming latent rank stability, the CiC identifies quantile treatment effects and other distributional contrasts, where the CiC transformation coincides with the monotone optimal transport (OT) map in one dimension.
A parallel literature has addressed the inadequacy of classical two-way fixed-effects estimators under staggered treatment adoption and effect heterogeneity, establishing that standard regression weights can be negative or difficult to interpret \citep[e.g.,][]{dCDH2020, GoodmanBacon2021}, which has led to the redefined target estimands at the cohort–time level and new estimators under more transparent causal contrasts \citep{CallawaySantAnna2021, SunAbraham2021}. Recently, most relevant to our work, by recasting distributional DiD within the geometry of OT, \citet{torous2024optimal} formulated causal identification as the minimization of a Monge-Kantorovich transportation cost for untreated outcome distributions over time,  generalizing both the additive translation of classical DiD and the monotone rearrangement of CiC via Brenier's theorem \citep{brenier1991polar}. A brief discussion on these related works can be found in the supplement. This OT-based paradigm unifies several existing models and provides a natural, rigorous foundation for nonlinear and multivariate settings. The use of transport distances for statistical inference has also generated a substantial statistics literature, including goodness-of-fit testing based on Wasserstein distances \citep{delbarrio1999tests}, distributional inference for empirical Wasserstein distances \citep{sommerfeld2018inference}, and broader statistical developments in Wasserstein space \citep{panaretos2019statistical}.

These developments signify a broader conceptual shift in modern DiD analysis: the object of interest is often best understood as the discrepancy between an observed post-treatment distribution and a counterfactual distribution generated by a structural evolution map. This perspective is inherently accommodated by the OT framework. Let \(\mu_0\) and \(\mu_1\) denote the pre- and post-period outcome distributions in the control group, respectively, and let \(d\) denote the map describing the natural untreated evolution from \(\mu_0\) to \(\mu_1\). In the univariate setting,
\(
d=F_{\mu_1}^{-1}\circ F_{\mu_0}
\)
corresponds to the monotone OT map pushing forward \(\mu_0\) to \(\mu_1\). Applying this map to the treated group’s pre-treatment law \(\mu_0^\ast\) yields the counterfactual post-treatment distribution
\(
\tilde\mu_1=d_\#\mu_0^\ast,
\)
that would have arisen in the absence of treatment. Thus, testing the null hypothesis of no treatment effect simplifies to evaluating the distributional equality:
\begin{equation}\label{dist.eq}
    H_0:\mu_1^\ast=\tilde\mu_1,
\end{equation}
where \(\mu_1^\ast\) represents the observed post-treatment law in the treated group. This formulation disentangles the natural temporal evolution learned of the control group and the treatment-induced deviation. However, while the OT-based approach identifies the counterfactual distribution, identification alone does not provide a formal inferential procedure. Operationalizing \eqref{dist.eq} as a global equality of two distributions remains an open question:
\begin{center}
\emph{Is it possible to construct a statistically principled test statistic, characterize its asymptotic null distribution, analyze its local power properties, and calibrate the test for finite samples?}
\end{center}

{Recently, \citet{linKongWang2023causalDistributionFunctions} and \citet{bhattacharjee2025doubly}
defined and estimated causal
effects for non-Euclidean outcomes in Wasserstein and metric spaces, respectively. \cite{tan2026unified} developed a mediation analysis framework that accommodates object-valued exposures, mediators, and outcomes. \cite{kurisuZhouOtsuMuller2024geodesicCausal} and \cite{zhouKurisuOtsuMuller2025geodesicDID} extended causal inference and DiD frameworks to non-Euclidean outcomes using geodesic formulations. Our approach offers a distinct, complementary perspective; specifically, we retain the OT-based counterfactual
structure for distributional DiD and focus on the development of a novel RKHS-based inferential test
for evaluating the equality between observed and transported counterfactual laws.
}

Our starting point is to interpret \eqref{dist.eq} as a problem of distributional equality within a reproducing kernel Hilbert space (RKHS). This connects the proposed causal testing problem to the kernel two-sample testing \citep{gretton2012kernel}. Specifically, we compare the observed post-treatment law \(\mu_1^\ast\) with the counterfactual law \(\tilde\mu_1=d_\#\mu_0^\ast\) via the maximum mean discrepancy (MMD), which is the squared RKHS distance between their respective kernel mean embeddings. This yields a fully nonparametric test for treatment heterogeneity under the OT-based DiD model. Because characteristic kernels metrize the topology of weak convergence at the embedding level, the resulting test is omnidirectional, possessing power against general departures from \eqref{dist.eq}, including mean shifts, scale changes, tail deformations, and other nonlinear distributional changes. This RKHS perspective allows us to make four distinct contributions that, taken together, distinguish the novelty of our paper.

First, we elevate the OT representation of distributional DiD from an identification device into a formal inferential framework. Existing OT-based formulations, such as \citet{torous2024optimal}, focus on constructing the counterfactual map but stop short of providing a general hypothesis test for evaluating whether the treated distribution deviates from that counterfactual law. Our approach closes this gap by embedding the causal identification strategy directly into an RKHS testing framework, adding a rigorous inferential layer to the OT-based causal framework. To our knowledge, this is the first RKHS-based inferential procedure for testing treatment heterogeneity under an OT DiD model. 

Second, unlike classical DiD and its extensions, we target the entire outcome distribution rather than a specific functional (such as the mean) or a finite collection of quantiles. Classical DiD fails to detect treatment heterogeneity whenever the mean effect is negligible, while CiC is constrained by a one-dimensional rank structure to capture distributional effects. By comparing probability measures directly, our framework is not limited to any pre-specified features of interest and provides a unified way to detect a broad class of treatment effects.

Third, we develop a rigorous asymptotic theory for the proposed MMD-based statistic. Under the null hypothesis, the statistic is governed by a second-order von-Mises expansion and converges to a weighted chi-square limit, or equivalently, the squared norm of a Gaussian random element in the RKHS. Furthermore, we characterize its behavior under contiguous local alternatives, thereby identifying the regime of nontrivial local power, and we establish the consistency result to guarantee asymptotic power against fixed alternatives.

Fourth, we provide a practical calibration scheme based on the spectral decomposition of empirical covariance operators. This delivers an implementable testing procedure that yields valid critical values and \(p\)-values, while accounting for the plug-in estimation of the transport map and associated nuisance quantities. Thus, our work bridges the theory and practical implementation by providing an inferential method that remains sensitive to general distributional departures without being confined to mean-level or quantile-level effects.

Taken together, these contributions advance the distributional DiD literature in two ways. Conceptually, they show that the OT-based counterfactual representation can seamlessly accommodate formal statistical inference alongside nonparametric identification. Methodologically, they provide an omnibus nonparametric test that is both flexible and theoretically grounded. In empirical applications where treatment may affect the outcome distribution in complex ways, this combination of transport geometry and RKHS-based inference provides a principled and powerful alternative to classical mean-based approaches.

The rest of this paper is organized as follows. Section~\ref{sec:problem} formulates the  OT-based DiD problem. Section~\ref{sec:method} develops the RKHS-based test statistic and its asymptotic theory. Section~\ref{sec:sim} presents the simulation results. Section~\ref{sec:data} presents the empirical application. Section~\ref{sec:concl} includes concluding remarks. More details and full proofs are presented in the supplement. 

\section{Problem Formulation}
\label{sec:problem}
We consider a standard two-group, two-period Difference-in-Differences (DiD) setting. Let
${D} \in \{0,1\}$, with the treated group {$D=1$}
and the control group ${D}=0$. Outcomes are observed at two time points $t \in \{0,1\}$: pre-intervention ($t=0$) and post-intervention ($t=1$).

For the control group, let $Y_t \sim \mu_t$, where $\mu_t$ denotes the marginal
distribution of the outcome at time $t=0,1$ with the joint law $(Y_0,Y_1) \sim G_0$. We assume that $\mu_0$ and $\mu_1$ arise as
push-forwards of a latent, time-invariant distribution $\nu$ through measurable production
functions $h_0$ and $h_1$, respectively. The latent distribution $\nu$ captures intrinsic
unit-level characteristics that remain stable over time.
For the treated group, let $Y_t^\ast \sim \mu_t^\ast$, where the marginal distribution $\mu_0^\ast$ is generated
from a latent distribution $\nu^\ast$ that may differ from $\nu$. The joint distribution of $(Y_0^\ast, Y_1^\ast)$ is denoted by $F_0$. Treatment is administered
only between $t=0$ and $t=1$, so that $\mu_1^\ast$ reflects both natural temporal evolution
and the effect of treatment. Outcomes are observed for both groups at both time points, but the
treatment effect is realized only for the treated group in the post-intervention period.

The outcome process evolves differently over time for the two groups. For the
control group, the evolution from $\mu_0$ to $\mu_1$ reflects a natural drift independent of
intervention. For the treated group, the post-intervention distribution $\mu_1^\ast$ combines
this natural drift with the treatment effect. Causal inference aims to disentangle these two
components by using the control group to identify the natural drift and applying it to the treated
group to generate the counterfactual distribution that would have been observed in the absence
of treatment.

Formally, 
the control group identifies a transport map $d : \mathbb{R} \to
\mathbb{R}$ satisfying $\mu_1 = d_{\#}\mu_0$, where $d_{\#}$ denotes the push-forward operator.
In the univariate setting, we take
$d = F_{\mu_1}^{-1} \circ F_{\mu_0}$,
the monotone OT map from $\mu_0$ to $\mu_1$. 
We assume that $d$ is also the push-forward map of the distributions of the treatment group if it were not treated. That is, $d$ represents the distributional natural drift in the treatment group.
This map is a
distributional analogue of the classical parallel trends assumption: it specifies how outcomes
would evolve over time in the absence of treatment at the level of entire distributions rather than only their means.

\begin{figure}[ht]
    \centering
    \includegraphics[width=.75\textwidth]{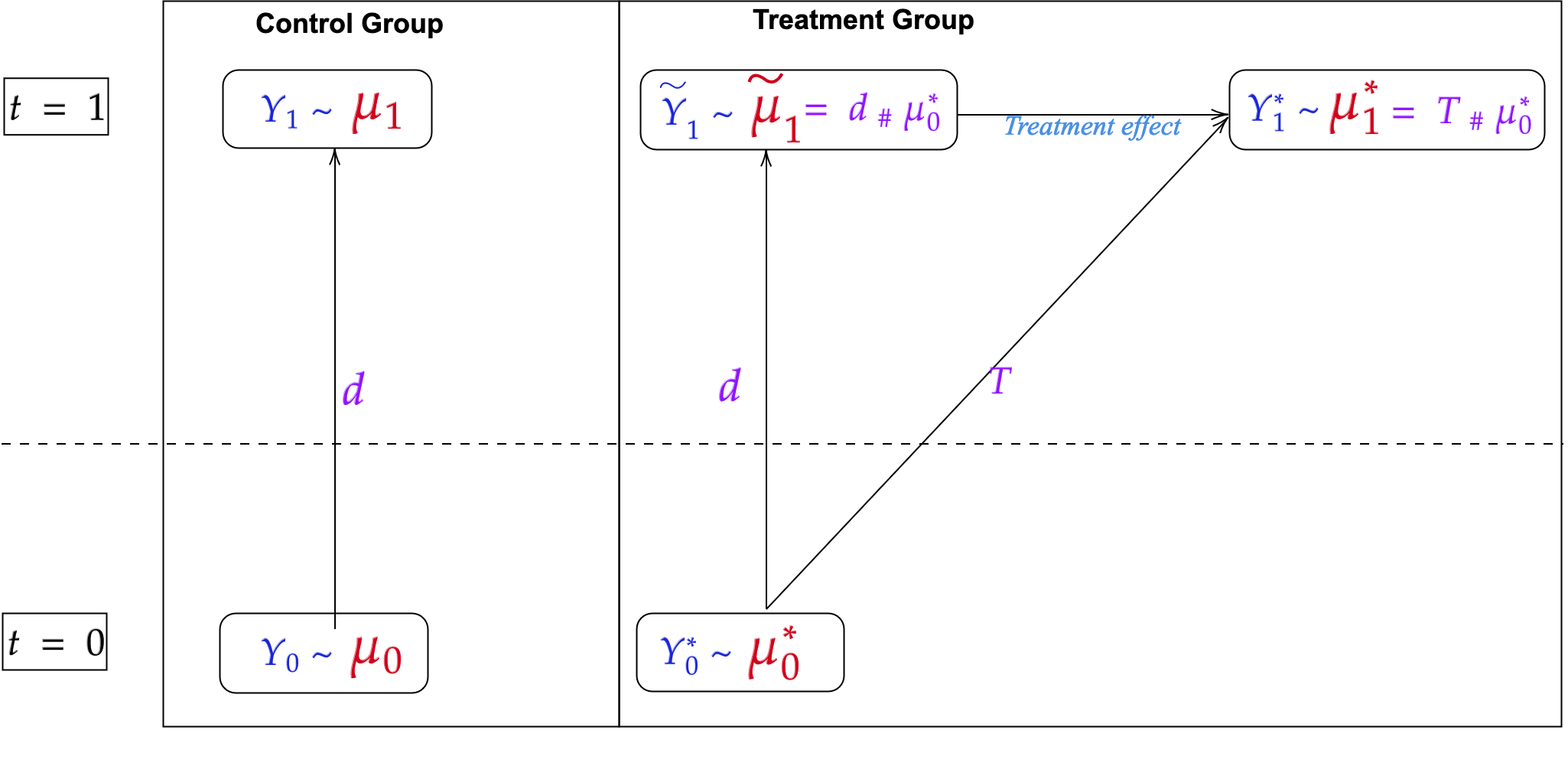}
    \caption{An illustration of OT maps in the space of measures and corresponding measurable random objects. 
    Arrows represent push-forward maps: $\mu_1 = d\#\mu_0$ and $\mu_1^\ast = T\#\mu_0^\ast$, with $d$ being the natural drift map common to both groups. The counterfactual distribution $\tilde{\mu}_1 = d\# \mu_0^\ast$ is observed if there is no treatment effect. 
    }
    \label{fig:OT:diagram}
\end{figure}

Figure~\ref{fig:OT:diagram} illustrates this structure. The random variables and their
distributions fall into five groups: (i) pre-intervention control outcomes $Y_0 \sim \mu_0$,
(ii) post-intervention control outcomes $Y_1 \sim \mu_1$, (iii) pre-intervention treated
outcomes $Y_0^\ast \sim \mu_0^\ast$, (iv) post-intervention treated outcomes
$Y_1^\ast \sim \mu_1^\ast$, and (v) post-intervention counterfactual treated outcomes
$\widetilde{Y}_1 \sim \widetilde{\mu}_1$, where $\widetilde{\mu}_1$ is unobserved. The
distributions $\mu_0$, $\mu_1$, $\mu_0^\ast$, and $\mu_1^\ast$ are observed, while
$\widetilde{\mu}_1$ represents the potential outcome distribution under no treatment. Expressed in terms of these symbols, our parallel-trends condition is rigorously formulated as the following assumption. 
\begin{assumptions}\label{eq:parallel trends}
\quad  If $d = F \lo {\tilde \mu \lo 1} \inv \circ F \lo {\mu \lo 0 \hi \ast}$, then $ \tilde \mu \lo 1 = d \lo {\#} \mu \lo 0 \hi *$.
\end{assumptions}

 Equivalently, the counterfactual outcomes satisfy $\widetilde{Y}_1 = d(Y_0^\ast)$. Another commonly used notation for $\tilde \mu \lo 1 = d \lo {\#} \mu \lo 0 \hi *$ is $\tilde \mu \lo 1 = \mu \lo  0 \hi * \circ d \inv$. In contrast,
the observed post-intervention distribution $\mu_1^\ast$ is generated from $\mu_0^\ast$ via
an unknown treatment effect map $T$, so that $\mu_1^\ast = T_{\#}\mu_0^\ast$.

We observe $m$ paired samples $(Y_0^j,Y_1^j)_{j=1}^m \stackrel{\text{i.i.d.}}{\sim} G_0$ from the
control group and $n$ paired samples $(Y_0^{\ast i},Y_1^{\ast i})_{i=1}^n
\stackrel{\text{i.i.d.}}{\sim} F_ 0$ from the treated group. Assuming these two samples are independent, their empirical measures 
$F_n$ and $G_m$ are also independent.
The drift map $d$ is estimated from the control group using the empirical distributions
$\hat{F}_{\mu_0}$ and $\hat{F}_{\mu_1}$ via
$\hat{d} = \hat{F}_{\mu_1}^{-1} \circ \hat{F}_{\mu_0}$. Under the null hypothesis of no treatment effect, the observed and
counterfactual post-intervention distributions for the treated group coincide. We therefore test:
\begin{align}
    \label{eq:null}
    H_0 : \widetilde{\mu}_1 = \mu_1^\ast
\quad \text{versus} \quad
H_1 : \widetilde{\mu}_1 \neq \mu_1^\ast.
\end{align}
Under $H_0$, the treated group evolves according to the same drift map as the control group,
and any deviation between $\mu_1^\ast$ and $\widetilde{\mu}_1$ constitutes evidence of
heterogeneous or nonlinear treatment effects. This formulation generalizes classical DiD by
replacing mean-level parallel trends with a distributional transport restriction, and reduces causal inference to a problem of distributional equality between the observed post-intervention law $\mu_1^\ast$ and the counterfactual law $\widetilde{\mu}_1 = d_{\#}\mu_0^\ast$. Since $\widetilde{\mu}_1$ is not observed but is an explicit functional of the joint law $(F_0,G_0)$, testing $H_0:\mu_1^\ast=\widetilde{\mu}_1$ naturally leads to the study of a statistical functional that compares $\mu_1^\ast$ with $d_{\#}\mu_0^\ast$, which will be developed in the subsequent sections.

\section{Test of Homogeneity}
\label{sec:method}
We construct a test based on the maximum mean discrepancy (MMD) defined as follows. 
\begin{definition}
Let $\mathcal{F}$ be a class of functions $f: \ca M \to \mathbb{R}$. The maximum mean discrepancy (MMD) between the observed and counterfactual outcome distributions is defined as
\[
\mathrm{MMD}(\mathcal{F}, \mu_1^\ast, \tilde{\mu}_1) := \sup_{f \in \mathcal{F}} \left(E_{\mu_1^\ast}[f(Y_1^{ \ast})] -E_{\tilde{\mu}_1}[f(\tilde{Y}_1)]\right).
\]
\end{definition}
We seek to identify a function class $\mathcal{F}$ that is rich enough to uniquely characterize the null hypothesis, while remaining sufficiently structured to enable reliable finite-sample estimation. One useful approach is to assume $\mathcal{F}$ as the unit ball in an RKHS and employ kernel mean embedding in the RKHS to measure the discrepancy between two distributions $\mu_1^\ast$ and $\tilde{\mu}_1$.

\subsection{The test statistic}
\label{sec:notation}
Let $(\Hc,\ip{\cdot}{\cdot})$ be a separable RKHS over $\mathbb R$ with reproducing kernel $\kappa:\mathbb R\times\mathbb R\to\mathbb R$.
Bochner integrals in $\Hc$ are used for expectations of $\Hc$–valued random elements. We assume that $\ka$ is bounded and continuous, i.e., there exists a $C > 0$, such that $\sup_{x\in \ca M} \ka(x, x') \leq C$.  By the definition of RKHS, the evaluation map $\delta \lo x: f \mapsto f(x)$ from $\ca H$ to $\real$ is continuous, implying that the Riesz representation of $\delta \lo x$ is $\ka (\cdot, x)$ with
$\delta \lo x (f) = f(x) = \langle f, \ka(\cdot, x) \rangle \lo{\ca H} $ for each $f \in \ca H$.

Define the kernel mean embedding of any random variable $Y_0^\ast$ following distribution $P$ as  $m_P \in \mathcal{H}$ such that $E_{Y_0^\ast\sim P}[f(Y_0^\ast)] = \langle f, m_P \rangle_\mathcal{H}$ for all $f \in \mathcal{H}$. 
If $k(\cdot, \cdot)$ is measurable and $E_{Y_0^\ast\sim P} \sqrt{k(Y_0^\ast, Y_0^\ast)} < \infty$, then $m_P \in \mathcal{H}$. For a characteristic kernel $\ka$, the mean embedding map $m_P$ is injective, and thus, distinct distributions map to distinct points in an RKHS. 
Further, the MMD can be written as the distance in $\mathcal{H}$ between embeddings $m_P$ and $m_Q$ as 
\begin{align}
    \label{eq:MMD:def}
\text{MMD}^2(\mathcal{H}, P, Q) = \|m_P - m_Q\|_\mathcal{H}^2.
\end{align}
This formulation follows the RKHS two-sample testing framework of \citet{gretton2012kernel}. 
Given that the mean embedding $m_P$ is injective, $\text{MMD}(\mathcal{H}, P, Q) = 0$ implies $P = Q$. This property holds for all RKHSs associated with characteristic kernels $\ka$~\citep{Sriperumbudur2010}, including, for example, the Gaussian and Laplace kernels.

The distributions of interest are the observed post-intervention law
$\mu_1^\ast$ and the counterfactual law $\widetilde\mu_1=d_{\#}\mu_0^\ast$. Adapting (\ref{eq:MMD:def}) to this setting, we have $\text{MMD}^2 ( \mathcal{H}, \mu_1^\ast, \tilde{\mu}_1) = \| m \lo  {\mu_1^\ast} - m \lo {\tilde{\mu}_1} \|\lo{\ca H} \hi 2 $. 
Although $\widetilde\mu_1$ is not observed, its RKHS embedding can be expressed using  
Assumption \ref{eq:parallel trends}:
$m_{\tilde{\mu}_1}  = \E_{\tilde{Y}_1 \sim \tilde{\mu}_1} \left[ \ka (\cdot, \tilde{Y}_1) \right] 
     =  \int \ka(\cdot, \tilde{Y}_1) d \tilde{\mu}_1 
    = \int \ka(\cdot, \tilde{Y}_1) d (d \lo  \# \mu_0^\ast) 
    = \E_{Y_0^\ast \sim \mu_0^\ast} \left[ \ka (\cdot, d(Y_0^\ast)) \right]$.
Thus, the population-level MMD between $\mu_1^\ast$ and $\widetilde\mu_1$ can be written as follows:
\begin{align}\label{eq:population mmd}
 \mathrm{MMD}^2({\ca H}, \mu_1^\ast,\widetilde\mu_1)
=\bigl\|
\mathbb E_{Y_1^\ast\sim\mu_1^\ast}\kappa(\cdot,Y_1^\ast)
-
\mathbb E_{Y_0^\ast\sim\mu_0^\ast}\kappa(\cdot,d(Y_0^\ast))
\bigr\|_{\mathcal H}^2,
\end{align}
which serves as the population target of the proposed test. This also quantifies the \emph{distributional treatment effect} (see the supplement for details). Its empirical estimate is
constructed by replacing expectations with averages and the drift map $d$ with its
estimator $\hat d=\theta(G_m)$ obtained from the control group. Expanding the quadratic form in (\ref{eq:population mmd}), we have
\begin{align}
    \label{eq:target:MMD}
\begin{split}
&   \text{MMD}^2 ({\mathcal{H}}, \mu_1^\ast, \tilde{\mu}_1) \\
 & = E_{Y_1^{\ast}, Y_1^{\ast '}} \left[ \ka (Y_1^{\ast}, Y_1^{\ast '}) \right] - 2E_{Y_1^{\ast}, \tilde{Y_1}} \left[ \ka (Y_1^{\ast}, \tilde{Y}_1) \right] +E_{\tilde{Y}_1, \tilde{Y}_1^{'}} \left[ \ka (\tilde{Y}_1, \tilde{Y}_1^{'} ) \right] \\
&  = E_{Y_1^{\ast}, Y_1^{\ast '}} \left[ \ka (Y_1^{\ast}, Y_1^{\ast '}) \right] - 2E_{Y_1^{\ast}, Y_0^\ast} \left[ \ka (Y_1^{\ast}, d(Y_0^\ast)) \right] +E_{Y_0^\ast, Y_0^{\ast '}} \left[ \ka (d(Y_0^\ast), d(Y_0^{\ast '})) \right],
\end{split}
\end{align}
where $Y_1^{\ast '}$ is an independent copy of $Y_1^{\ast}$ with the same distribution $\mu_1^\ast$, and $\tilde{Y}_1^{'}$ is an independent copy of $\tilde{Y}_1$ with the distribution $\tilde{\mu}_1$.
Testing $H_0$ in~\eqref{eq:null} thus reduces to testing 
\begin{align}
    \label{eq:null2}
    H_0: \text{MMD}^2 ({\mathcal{H}}, \mu_1^\ast, \tilde{\mu}_1) =0 \text{\ \  versus \ } H_1: \text{MMD}^2 ({\mathcal{H}}, \mu_1^\ast, \tilde{\mu}_1) >0.
\end{align} 
Define the symmetric kernel
$h_{n,m}(i,j)
:=
\kappa(Y^{\ast i}_1,Y^{\ast j}_1)
+\kappa(\hat d(Y^{\ast i}_0),\hat d(Y^{\ast j}_0))
-\kappa(Y^{\ast i}_1,\hat d(Y^{\ast j}_0))
-\kappa(Y^{\ast j}_1,\hat d(Y^{\ast i}_0))$,
where $h$ is indexed by $(m,n)$ as $\hat d $ is calculated from the control sample of $m$ subjects. 
The V-statistic  corresponding to the population-level $\mathrm{MMD} \hi 2$ in~\eqref{eq:target:MMD} is defined as
\begin{align}
    \label{eq:V-stat0}
 V_{n,m}^2 = & \frac{1}{n^2}\sum_{i=1}^n\sum_{j=1}^n h_{n,m}(i,j).
\end{align}  
$V_{n,m}^2$ measures the discrepancy between the empirical distribution of the post-period outcomes \(\{Y_{1,i}^\ast\}\) and that of the transported pre-intervention outcomes \(\{\hat{d}(Y_{0,i}^\ast)\}\) in the RKHS induced by \(\kappa\). Under $H_0$, these two distributions coincide, meaning $V_{n,m}^2$ should be small.
This formulation is intrinsic to the problem, arising directly from comparing the two empirical RKHS embeddings to align with the population discrepancy of interest. The asymptotic behavior of $V_{n,m}^2$ is governed jointly by the fluctuations of the empirical averages within the definition of \(h_{n,m}\) and the estimation error of the transport map \(\hat d\).

An alternative test statistic can be constructed from the corresponding U-statistic:
     $U_{n,m}^2 = \frac{1}{n(n-1)}\sum_{i=1}^n \ \, \sum_{\substack{j=1,  j\ne i}}^n h_{n,m}(i,j)$.
If $d$ were known, this U-statistic would be an unbiased estimator of the population-level $\mathrm{MMD} \hi 2 ( \ca H, \mu \lo 1 \hi *, \tilde \mu \lo 1)$. In our setting, however, it is biased since it involves the estimated OT maps; thus, the typical advantage of a centered U-statistic is largely absent. Conversely, a test statistic based on the V-statistic serves as the natural plug-in estimator for the functional Von-Mises expansion~\citep{fernholz1983vonmises}, as detailed in Section~\ref{sec:von-mises}. Thus, we focus on the V-statistic $V_{n,m}$ throughout the remainder of our paper.

We will analyze the squared MMD as a real-valued functional of the joint laws $F_0$ of $(Y_0^\ast,Y_1^\ast)$ and $G_0$ of $(Y_0,Y_1)$:
\(
T(F_0,G_0)  = \text{MMD}^2 (\mathcal{H}, \mu_1^\ast, d_\#\mu_0^\ast ).
\)
This formulation allows us to study the asymptotic behavior via
von-Mises expansions along empirical paths $(F_n,G_m)$.

 \subsection{Von-Mises expansion of statistical functional}
 \label{sec:von-mises}

Define the linear paths $F_\epsilon:=(1-\epsilon)F_0+\epsilon F_n$ and $G_\eta:=(1-\eta)G_0+\eta G_m$ for $(\epsilon,\eta)\in[0,1]^2.$ To facilitate von-Mises expansion, we introduce the functional, $\theta(\cdot)$, for the  OT map: 
$\theta(G_0):=d = F_{\mu_1}^{-1}\circ F_{\mu_0}:\mathbb R\to\mathbb R$,
where $F_{\mu_0},F_{\mu_1}$ are the marginals of $G_0$ and $F_{\mu_1}^{-1}$ is the (left–continuous) quantile function. In this notation, $\theta (G \lo 0)$ is the pushforward transformation $d$, and $\theta (G \lo m)$ is its estimate $\hat d$: 
 $\theta (G \lo m) :=  \hat d = \hat F \lo {\mu \lo 1, m} \inv \circ \hat F \lo {\mu \lo 0, m}$, 
where $\hat F \lo {\mu \lo 1, m}$ is the empirical distribution based on the $m$ i.i.d. observations on $Y \lo 1$, and $\hat F \lo {\mu \lo 0, m}$ is the empirical distribution based on the $m$ i.i.d. observations on $Y \lo 0$. {The statistical OT is also an active research area \citep{hutter2021minimax_ot_maps,deb2021rates_ot_maps,
manole2024plugin_ot_maps,ding2024statistical,li2026stability,zhang2026copula}.}
The V-statistic in~\eqref{eq:V-stat0} can be rewritten as a functional of $F_n$ and $G_m$:
\begin{align}
    \label{eq:V-stat}
  V_{n,m}^2 = 
\ \left\|
\ \mathbb{E}_{F_n}\big[\kappa(\cdot,\theta(G_m)(Y_0^\ast))\big]
-\mathbb{E}_{F_n}\big[\kappa(\cdot,Y_1^{ \ast})\big]\ \right\|_{\Hc}^{2} 
:=   T(F_n, G_m).  
\end{align}
For $(\epsilon,\eta)\in[0,1]^2$ define
$\tau(\epsilon,\eta) := \left\|
 \mathbb{E}_{F_\epsilon}\big[\kappa(\cdot,\theta(G_\eta)(Y_0^\ast))\big]
-\mathbb{E}_{F_\epsilon}\big[\kappa(\cdot,Y_1^{ \ast})\big] \right\|_{\Hc}^{2}\ \in\ \mathbb R_+ $.
Clearly, $V_{n,m}^2 =  T(F_n, G_m) = \tau(1,1)$, and $T(F_0, G_0) = \tau(0,0)$. The latter equals $0$ under the null.

We will apply the von-Mises expansion to $V_{n,m}^2$ in~\eqref{eq:V-stat}, as it is a direct function of the joint empirical distributions $F \lo n$ and $G \lo m$. To develop the asymptotic distribution of the test statistic, we carry out a second-order von Mises expansion of the statistical functional $\tau(\cdot, \cdot)$ in a Banach space with a stochastic remainder term of negligible order. {We base our asymptotic inference on a scaled version of the plug-in empirical $\text{MMD}^2$ between the transported control distribution and the treated distribution, given by
\begin{align}
    \label{eq:test:stat:V_scaled}
    S_{n,m} &= 2\rho_{n,m}\, V_{n,m}^2 
    \text{ \  where \ \ } \rho_{n,m}:=\frac{nm}{n+m}.
\end{align}
This is the natural and appropriately scaled quadratic functional associated with the plug-in discrepancy $\tau(1,1)=T(F_n,G_m)$, which has a tractable asymptotic distribution. This choice of the scaled V-statistic is 
dictated by the structure of the functional $T(F,G)$: a second-order von Mises expansion around $(F_0,G_0)$ shows that, under $H_0:\tau(0,0)=0$, the first-order term vanishes and $S_{n,m}$ captures the leading nondegenerate fluctuation. Consequently, $S_{n,m}$ admits a weighted chi-square limit without further centering, providing a canonical pivot for testing. 
Moreover, since $V_n^2$ is a nonnegative estimator of the RKHS discrepancy, larger values of $S_{n,m}$ correspond to larger deviations from $\tau(0,0)=0$ and thus provide evidence against the null. Thus, $S_{n,m}$ is not merely convenient but intrinsic: it is the unique quadratic statistic that both reflects the geometry of $\tau$ and yields a directly calibratable null distribution.} 

Let \(\ca D\) denote the set of bivariate distribution functions on \(\mathbb R^2\), viewed as a subset of \(L^\infty(\mathbb R^2)\), the Banach space of bounded real-valued functions on \(\mathbb R^2\) equipped with the supremum norm
$
\|h\|_\infty := \sup_{(x,y)\in\mathbb R^2} |h(x,y)|.
$
We use $\ca D$ to denote both the family of control-group distributions $G$ and the family of treated-group distributions $F$.
Although every distribution function is uniformly bounded by \(1\), the supremum norm is used here to measure perturbations such as \(F_n-F_0\) and \(G_m-G_0\), which determine the stochastic order of the von Mises remainder and the differentiability properties of the transport functional. 

We make the following regularity assumptions.

\begin{assumptions}\label{ass:kernel}
The kernel $\kappa$ satisfies
$
\sup_{y\in\mathbb R}\kappa(y,y)\le K<\infty.
$
Moreover, for $r=0,1,2,3$ and each $y\in\mathbb R$, 
$\partial_2^r\kappa(\cdot,y)\in\mathcal H$, and the map
$y\mapsto \partial_2^r\kappa(\cdot,y)$ is continuous and uniformly bounded
in $\mathcal H$-norm, {where $\partial_2^r \kappa(\cdot,y)$ denotes the $r$-th derivative of $\kappa(x,y)$ with respect to  $y$.} 
\end{assumptions}

\begin{assumptions}
\label{ass:marginal} The marginal $F_{\mu_1}$ of $G_0$ admits a density $p_1$ that is bounded away from $0$ and has bounded, continuous derivatives up to order $3$ on an open interval containing the {support  of the distribution of } 
$\theta(G_0)(Y_0^\ast)$ when $(Y_0^\ast,Y_1^{ \ast})\sim F_0$.
\end{assumptions}

\begin{assumptions}
\label{ass:differentiable}
The map $\theta:\ca D\to L \hi \infty(\mathbb R),\ 
\theta(G)=F_{\mu_1}^{-1}\circ F_{\mu_0},$
is twice Hadamard differentiable at every \(G\) in a neighborhood of \(G_0\), tangentially to a linear subspace \(\mathcal T\subset L \hi \infty(\mathbb R^2)\) such that the sample paths 
$G_m - G_0 \in \mathcal T$ almost surely for each $m.$
The corresponding first and second derivatives, $D\theta(G)$ and $D^2\theta(G)$,
are continuous as a linear and bilinear map, respectively. Moreover, there exist constants \(\delta>0\) and \(L<\infty\) such that
$\|D^2\theta(G_1)-D^2\theta(G_2)\|_{\mathrm{BL}(L \hi \infty(\mathbb R^2) \times L \hi \infty(\mathbb R^2), L \hi \infty(\mathbb R))}
\le L\|G_1-G_2\|_\infty$,
whenever \(\|G_1-G_0\|_\infty\le \delta\) and \(\|G_2-G_0\|_\infty\le \delta\). Here, \(\mathrm{BL}(S\times T, T)\) denotes the space of bounded bilinear maps from \(S\times S\) into \(T\), equipped with 
$
\|B\|_{\mathrm{op}}=\sup_{\|s_1\|_S\le1,\ \|s_2\|_S\le1}\|B(s_1,s_2)\|_T.
$
\end{assumptions}

On \(L \hi \infty(\mathbb R^2)\times L \hi \infty(\mathbb R^2)\), we use the product norm
$\|(h_1,h_2)\|_{\oplus} := \|h_1\|_\infty + \|h_2\|_\infty.$ Alternatively, one could employ $\max\{ \|h_1\|_\infty, \|h_2\|_\infty \}$, which is topologically equivalent. 
\begin{assumptions}
    \label{ass:bochner} $\Hc$ is separable, and the Bochner map
\(
M(F,G):=\mathbb{E}_F\big[\kappa(\cdot,\theta(G)(Y_0^\ast))\big]-\mathbb{E}_F\big[\kappa(\cdot,Y_1^{ \ast})\big]\in\Hc
\) from $\big(\ca D \times \ca D ,\,\|\cdot\|_\oplus \big)$ to $(\Hc,\|\cdot\|_{\Hc})$
is three times Fr\'echet differentiable in a neighborhood of $(F_0,G_0)$, with a bounded third Fr\'echet derivative on this neighborhood. 
\end{assumptions}
Note that the discrepancy functional satisfies
$
T(F,G) = \|M(F,G)\|_{\mathcal H}^2,
$
so that $M(F,G)$ is the underlying RKHS-valued moment whose squared norm defines the test statistic.

Assumptions \ref{ass:kernel}--\ref{ass:bochner} are standard in second-order
asymptotic analysis of nonparametric and semiparametric functionals involving estimated nuisance
components. The detailed justification of these assumptions is presented in the supplement. 

Since $(u\mapsto \|u\|_{\Hc}^{2})$ is three times Fr\'echet differentiable with bounded derivatives on $\Hc$, using the composition theorem, we obtain the following lemma on the smoothness of $\tau$:

\begin{lemma}[Smoothness of $\tau$]
\label{lem:smooth}
Under Assumptions \ref{ass:kernel}–\ref{ass:differentiable}, 
the map $\tau: \ca D \times \ca D \to\mathbb R$ is three times Fr\'echet differentiable in a neighborhood of $(F_0,G_0)$, with a bounded third Fr\'echet derivative on that neighborhood. 
\end{lemma}

\noindent With Lemma \ref{lem:smooth}, we may invoke the multivariate Taylor theorem in Banach spaces:

\begin{lemma}[Taylor's theorem with integral remainder]\label{lem:taylor}
Let $B$ be a Banach space, $U\subset B$ open, $b_0\in U$, and $\Phi:U\to\mathbb R$ three times Fr\'echet differentiable on the line segment $\{b_0+th: t\in[0,1]\}\subset U$ with bounded third derivative there. Then, for every $h$ with $b_0+h\in U$,
\[
\Phi(b_0+h) = \Phi(b_0) + D\Phi[b_0]\{h\} +\tfrac{1}{2}D^2\Phi[b_0]\{h,h\} + R_3(b_0,h),
\]
where 
\(
R_3(b_0,h)=\int_{0}^{1}\tfrac{(1-t)^2}{2}\ D^3\Phi[b_0+th]\{h,h,h\}\,dt
\)
is the integral remainder satisfying that
$$|R_3(b_0,h)|\ \le\ \tfrac{1}{6}\ \sup_{t\in[0,1]}\|D^3\Phi[b_0+th]\|\ \cdot \ \|h\|^{3}.
$$
\end{lemma}
The proof of Lemma \ref{lem:taylor} can be found in~\cite{zeidler2012applied} and is thus omitted.

\begin{proposition}[Von-Mises expansion in Banach spaces]\label{prop:von-mises}
Under Assumptions~\ref{ass:kernel}--\ref{ass:bochner}, the functional $\tau$ admits the following second-order Von-Mises expansion:
\begin{align}
\tau(1,1)
&=\tau(0,0)\ +\ \left.\frac{\partial}{\partial \epsilon}\tau(\epsilon,0)\right|_{\epsilon=0}\,\ +\ \left.\frac{\partial}{\partial \eta}\tau(0,\eta)\right|_{\eta=0}\,\nonumber\\
&\quad+\ \frac{1}{2}\left.\frac{\partial^2}{\partial \epsilon^2}\tau(\epsilon,0)\right|_{\epsilon=0}\,\ +\ \frac{1}{2}\left.\frac{\partial^2}{\partial \eta^2}\tau(0,\eta)\right|_{\eta=0}\,\ +\ \left.\frac{\partial^2}{\partial \epsilon\,\partial \eta}\tau(\epsilon,\eta)\right|_{(\epsilon,\eta)=(0,0)} + R_{n,m}.\label{eq:main-expansion}
\end{align}
Under the null hypothesis~\eqref{eq:null3}, $\tau(0,0)=0$ and 
$
 \left. \frac{ \partial \tau (\epsilon, 0) }{\partial \epsilon } \right| \lo {\epsilon = 0} = 0, \quad   \left. \frac{ \partial \tau (0, \eta) }{\partial \eta } \right| \lo {\eta = 0} = 0.
$
Furthermore, the stochastic remainder obeys
$$R_{n,m}=O_P\!\big((\|F_n-F_0\|_\infty+\|G_m-G_0\|_\infty)^3\big)=O_P(N^{-3/2}),$$
whenever $\|F_n-F_0\|_\infty=O_P(n^{-1/2})$ and $\|G_m-G_0\|_\infty=O_P(m^{-1/2})$ and $N\asymp n\asymp m$. 
\end{proposition}
{Here, \(n\asymp m\) denotes that \(n/m\) remains bounded away from both \(0\) and \(\infty\). The rate \(O_{\mathbb P}(N^{-3/2})\) follows from the bounded third derivative in the Banach-space Taylor expansion together with the Dvoretzky–Kiefer–Wolfowitz inequality, which implies the uniform empirical process bounds \(\|F_n-F_0\|_\infty=O_{\mathbb P}(n^{-1/2})\) and \(\|G_m-G_0\|_\infty=O_{\mathbb P}(m^{-1/2})\).}
The explicit expressions of the derivatives can be found in the supplement.

{Proposition~\ref{prop:von-mises} shows that the first-order derivative of \(T(F,G)\) along the empirical paths vanishes under $H_0$, so the asymptotic behavior is governed by the second-order term in the von-Mises expansion. The statistic thus admits a quadratic approximation in the empirical process. Combined with the weak convergence of the empirical process to a Gaussian limit, this yields convergence in distribution to a quadratic form in a Gaussian element of \(\mathcal H\), or equivalently, a weighted sum of independent squared Gaussian variables whose weights are the eigenvalues of the associated covariance operator. This second-order delta-method, or Wilks-type phenomenon (see, e.g., \cite{li2019graduate}), leads to a chi-square-type asymptotic distribution for the scaled statistic. We compute these derivatives in~\eqref{eq:main-expansion} and derive the resulting limit for \(V_n^2\), with full details deferred to the supplement.}

\subsection{Asymptotic distribution under the null hypothesis}
\label{sec:dist:null}

We now leverage the second-order expansion to characterize the limiting null distribution of the test statistic.  We first express $H_0$ in~\eqref{eq:null2}  as an equality in law between the transformed pre-treatment outcomes and the observed post-treatment outcomes in the treated group:
\begin{align}\label{eq:null3}
\mathcal L(Y_1^{ \ast}) =\mathcal L\big(\theta(G_0)(Y_0^\ast)\big)\quad\text{when} \quad (Y_0^\ast,Y_1^{ \ast})\sim F_0.
\end{align}
Here, $\mathcal L (X)$ denotes the distribution of any random variable $X$. Since $\theta(G_0)(Y_0^\ast)$ and $Y_1^\ast$ have the same distribution under $F_0$, their kernel mean embeddings in $\mathcal H$ coincide, yielding 
\begin{align}\label{eq:baseline-0}
\mathbb E_{F_0}\!\big[\kappa(\cdot,\theta(G_0)(Y_0^\ast))\big]
-\mathbb E_{F_0}\!\big[\kappa(\cdot,Y_1^{ \ast})\big]=0\quad\text{in }\Hc.
\end{align}
Equation~\eqref{eq:baseline-0} identifies $H_0$ as a population-level degeneracy
condition for the RKHS-valued moment map underlying the statistic. When the
empirical laws $(F_n,G_m)$ fluctuate around $(F_0,G_0)$, the leading contribution to the statistic
arises from second-order stochastic terms rather than from a linear expansion. The next result
formalizes this observation by characterizing the resulting limiting distribution under $H_0$. This is built directly on the reformulation of the null hypothesis and the second-order
von-Mises expansion established {earlier}. 

Under Assumption~\ref{ass:kernel}, for each $y\in\mathbb R$,  the map $y\mapsto \kappa(\cdot,y)\in\mathcal H$ is continuously
Fr\'echet differentiable with derivative $ \frac{d}{dy}\kappa(\cdot,y)=\partial_2\kappa(\cdot,y)\in\mathcal H.$
Consequently, every $h\in\mathcal H$ is continuously differentiable and satisfies
the derivative reproducing property
$
h'(y)
=
\langle h,\partial_2\kappa(\cdot,y)\rangle_{\mathcal H},
\ y\in\mathbb R.
$
We will use this identity to interpret derivatives of RKHS elements.
We also denote by $p_1$ the density of the control post-period marginal
distribution $F_{\mu_1}$ from Assumption~\ref{ass:marginal}.

\begin{theorem}[Asymptotic null distribution]\label{thm:main} 
Under Assumptions \ref{eq:parallel trends}--\ref{ass:bochner}, if
$\gamma_{n,m}:=\frac{n}{n+m}\to\gamma\in(0,1)$, then under the null hypothesis \eqref{eq:null3}, the test statistic 
$\mathcal S_{n,m}$ defined in~\eqref{eq:test:stat:V_scaled} satisfies
\begin{equation}
    \mathcal S_{n,m}\ \cid\ 2\sum_{k=1}^{\infty}\lambda_k\,\chi^2_{1,k},
\end{equation}
where $\{\lambda_k\}$ are the eigenvalues of the covariance operator 
$\mathcal C_\gamma=\gamma\,\mathcal C^{(G)}+(1-\gamma)\,\mathcal C^{(F)}$  on $\mathcal H$. More explicitly, for $f,g\in\mathcal H$, we have 
\begin{align*}
\langle f,\mathcal C_\gamma g\rangle_{\mathcal H}
&=(1-\gamma)\,
\mathbb E_{F_0}\!\left[
\bigl(f(\theta(G_0)(Y_0^\ast))-f(Y_1^\ast)\bigr)
\bigl(g(\theta(G_0)(Y_0^\ast))-g(Y_1^\ast)\bigr)
\right]\\
&\quad
+\gamma\,\mathbb E_{G_0}\!\left[
\frac{f'(\theta(G_0)(Y_0^\ast))}{p_1(\theta(G_0)(Y_0^\ast))}
\,
\frac{g'(\theta(G_0)(Y_0^{\ast'}))}{p_1(\theta(G_0)(Y_0^{\ast'}))}
\times\,
\Gamma\!\bigl((Y_0^\ast,\theta(G_0)(Y_0^\ast)),
(Y_0^{\ast'},\theta(G_0)(Y_0^{\ast'}))\bigr)
\right],
\end{align*}
where the covariance kernel $\Gamma$ is self-adjoint, positive, trace-class on $\mathcal H$, and is defined as follows: $\Gamma \big((x,y),(x',y')\big)
:=
F_{\mu_0}(\min\{x,x'\})-F_{\mu_0}(x)F_{\mu_0}(x')
 +F_{\mu_1}(\min\{y,y'\})-F_{\mu_1}(y)F_{\mu_1}(y')-G_0(x,y')-G_0(x',y)
+F_{\mu_0}(x)F_{\mu_1}(y')
+F_{\mu_0}(x')F_{\mu_1}(y)$.
\end{theorem}
Theorem~\ref{thm:main} shows that the null distribution of the scaled statistic is a weighted chi-square mixture, or equivalently, the squared norm of a Gaussian random element in \(\mathcal H\). The eigenvalues of \(\mathcal C_\gamma\) determine the contribution of each orthogonal RKHS direction to the null variability and therefore govern both critical values and asymptotic p-values.

The limiting distribution can alternatively be interpreted as a Gaussian quadratic form:
\begin{equation}\label{eq:chisq-mixture}
2\,\|Z_\gamma\|_{\Hc}^{2}\ \stackrel{\mathcal D}{=}\ 2\sum_{k=1}^{\infty}\lambda_k\,\chi^2_{1,k},
\end{equation}
where $Z_\gamma$ 
 is a mean-zero Gaussian random element in the separable
Hilbert space $\mathcal H$ with covariance operator $\mathcal C_\gamma$, and the Karhunen-Lo\`eve expansion yields
$
Z_\gamma=\sum_{k\ge1}\sqrt{\lambda_k}\,\xi_k e_k,
$
where $\{e_k\}$ is an orthonormal eigenbasis of $\mathcal C_\gamma$ and
$\{\xi_k\}_{k\ge1}\stackrel{\text{i.i.d.}}{\sim} N(0,1)$ 
\citep[Chapter 3]{vdVWellner1996}. The spectrum thus determines how different
orthogonal directions in the RKHS contribute to variability under the null. Consequently, critical values and $p$-values are defined through the quantiles and tail probabilities of the limiting distribution.

\paragraph{Asymptotic size-$\alpha$ test.}

Let $c_{1-\alpha}$ denote the $(1-\alpha)$ quantile of the limiting distribution in
\eqref{eq:chisq-mixture}. Because $\mathcal S_{n,m}$ converges in distribution to
$2\|Z_\gamma\|_{\Hc}^2$ under $H_0$ and the limiting mixture of $\chi \lo 1 \hi 2$ distribution is continuous at $c_{1-\alpha}$, the continuous mapping theorem implies that
$\varphi_{\alpha}=\mathbbm 1\{\ \mathcal S_{n,m}>c_{1-\alpha}\ \}$
has asymptotic size $\alpha$, i.e.,
$\lim_{n,m\to\infty}\mathbb P_0(\varphi_\alpha=1)=\alpha$ 
\citep[Section~2.8]{vdVWellner1996}. This continuity condition excludes point masses at the critical
value and ensures the stability of rejection probabilities under small perturbations.

\paragraph{Asymptotic $p$-value.}

For an observed $s\ge0$, the (oracle) $p$-value is defined as the upper-tail probability of
the exact limiting null distribution,
$p(s;\{\lambda_k\})
:=\mathbb P\!\left(2\sum_{k=1}^{\infty}\lambda_k\,\chi^2_{1,k}\ \ge s\right).
$ 
In practice, this distribution is approximated by a finite truncation
$Q_r:=2\sum_{k=1}^{r}\lambda_k\,\chi^2_{1,k}$. The cumulative distribution function of $Q_r$ admits
the Imhof inversion formula \citep{Imhof1961}:
\begin{equation}\label{eq:Imhof}
\mathbb P(Q_r\le s)
=\frac{1}{2}+\frac{1}{\pi}\int_{0}^{\infty}
\frac{\Im\!\left\{\exp\!\left(it\frac{s}{2}\right)
\prod_{k=1}^{r}(1-2it\lambda_k)^{-1/2}\right\}}{t}\;dt,
\end{equation}
obtained by Fourier inversion of the characteristic function of $Q_r$. The truncation error is controlled by Markov's inequality: 
$\mathbb P\!\left(2\sum_{k>r}\lambda_k\,\chi^2_1 \ge s\right)
\le \frac{2}{s}\mathbb E\big[\sum_{k>r}\lambda_k\,\chi^2_1\big]
= \frac{2}{s}\sum_{k>r}\lambda_k$. 
Since $\mathcal C_\gamma$ is trace-class, $\sum_k\lambda_k=\mathrm{tr}(\mathcal C_\gamma)<\infty$,
and hence the truncation error vanishes as $r\to\infty$. This guarantees that the oracle $p$-value
can be approximated arbitrarily well by finite-dimensional truncations. The oracle $p$-value quantifies extremeness relative to the exact null limit law, while the Imhof
representation provides a numerically exact method for evaluating the distribution of finite
truncations. The trace-class property of $\mathcal C_\gamma$ guarantees that the truncation error is
controlled and vanishes as additional eigencomponents are included.

\subsection{Empirical covariance operators and estimated eigenvalues}
\label{sec:est:null}
{We now describe the construction of the empirical covariance operator and the estimated eigenvalues used for calibration. Let \(\hat\theta:=\theta(G_m)\) denote the plug-in transport map estimated from the control sample; thus, in our context \(\hat\theta(y)=\hat F_{\mu_1}^{-1}(\hat F_{\mu_0}(y))\) for \(y\in\mathbb R\).}

{Let $\hat p_1$ be a strictly positive estimator of $p_1$ on the range of $\hat\theta(Y_0^{\ast i})$. Define the RKHS-valued quantities: $\hat \xi_i:=
\kappa(\cdot,\hat \theta(Y_0^{\ast i}))
-
\kappa(\cdot,Y_1^{\ast i})$,  $\hat w_j(y)
:=
\frac{
\mathbbm 1\{Y_0^j\le y\}-\hat F_{\mu_0}(y)
-
\big(\mathbbm 1\{Y_1^j\le \hat \theta(y)\}-\hat F_{\mu_1}(\hat \theta(y))\big)
}{
\hat p_1(\hat \theta(y))
},$ $\hat \zeta_j
:=
\frac{1}{n}\sum_{i=1}^n
\hat w_j(Y_0^{\ast i})\,
\partial_2\kappa(\cdot,\hat \theta(Y_0^{\ast i}))$, $\overline{\xi}
:=
\frac{1}{n}\sum_{i=1}^n \hat \xi_i$, and $\overline{\zeta}
:=
\frac{1}{m}\sum_{j=1}^m \hat \zeta_j$. 
Here, $\hat \xi_i$ represents the empirical RKHS discrepancy between the observed post-period outcome and its transported pre-period counterpart, while $\hat \zeta_j$ captures the contribution of the control sample through the linearization of the transport map. Leave-one-out versions of $\hat F_{\mu_0}$, $\hat F_{\mu_1}$, and $\hat \theta$ may be used to reduce finite-sample bias. The empirical covariance operator is given by
\begin{equation}
\label{eq:Cgamma}
\hat{\mathcal C}_\gamma
:=
(1-\hat\gamma)\frac{1}{n}\sum_{i=1}^n
(\hat \xi_i-\overline{\xi})\otimes(\hat \xi_i-\overline{\xi})
+
\hat\gamma\,\frac{1}{m}\sum_{j=1}^m
(\hat \zeta_j-\overline{\zeta})\otimes(\hat \zeta_j-\overline{\zeta})
\end{equation}
with $\hat\gamma=\frac{n}{n+m}$, 
which is a finite-rank, positive, self-adjoint operator on $\mathcal H$ aggregating variability from both samples.
To compute its spectrum, we luse a Gram representation. Let $H_n:=I_n-\frac{1}{n}\mathbf 1\mathbf 1^\top$ and $H_m:=I_m-\frac{1}{m}\mathbf 1\mathbf 1^\top$. Using the reproducing property, $\hat K^{(F)}_{i\ell}
=
\langle \hat \xi_i,\hat \xi_\ell\rangle_{\mathcal H}
=
\kappa(\hat \theta(Y_0^{\ast i}),\hat \theta(Y_0^{\ast \ell}))
-\kappa(\hat \theta(Y_0^{\ast i}),Y_1^{\ast \ell})
-\kappa(Y_1^{\ast i},\hat \theta(Y_0^{\ast \ell}))
+\kappa(Y_1^{\ast i},Y_1^{\ast \ell})$, 
and set $\widetilde K^{(F)}:=H_n \hat K^{(F)} H_n$. By the derivative reproducing property, $\hat M_{i\ell}=\partial_{22}\kappa(\hat \theta(Y_0^{\ast i}),\hat \theta(Y_0^{\ast \ell}))$. With $\hat W_{j,i}=\hat w_j(Y_0^{\ast i})$, define $\hat L^{(G)}:=\frac{1}{n^2}\hat W\,\hat M\,\hat W^\top$ and $\widetilde L^{(G)}:=H_m\,\hat L^{(G)}\,H_m$. The cross Gram matrix $\hat C^{(FG)}\in\mathbb R^{n\times m}$ has entries
$\langle \hat \xi_i,\hat \zeta_j\rangle_{\mathcal H}
=
\frac{1}{n}\sum_{\ell=1}^n \hat w_j(Y_0^{\ast \ell})
\Big(
\partial_2\kappa(\hat \theta(Y_0^{\ast i}),\hat \theta(Y_0^{\ast \ell}))
-
\partial_2\kappa(Y_1^{\ast i},\hat \theta(Y_0^{\ast \ell}))
\Big)$, and we set $\widetilde C^{(FG)}:=H_n\,\hat C^{(FG)}\,H_m$.
Writing $\hat{\mathcal C}_\gamma$ as a sum of rank-one operators based on centered and scaled feature vectors $v_i^{(F)}:=\sqrt{\frac{1-\hat\gamma}{n}}(\hat \xi_i-\overline{\xi})$ and $v_j^{(G)}:=\sqrt{\frac{\hat\gamma}{m}}(\hat \zeta_j-\overline{\zeta})$, its nonzero eigenvalues $\hat\lambda_1\ge \hat\lambda_2\ge\cdots\ge 0$ are used for calibration of the test statistic and coincide with those of the block Gram matrix 
\begin{equation}
\label{eq:Gram}
\hat G
=
\begin{bmatrix}
\frac{1-\hat\gamma}{n}\,\widetilde K^{(F)}
&
\sqrt{\frac{(1-\hat\gamma)\hat\gamma}{nm}}\,\widetilde C^{(FG)}
\\
\sqrt{\frac{(1-\hat\gamma)\hat\gamma}{nm}}\,\widetilde C^{(FG)\top}
&
\frac{\hat\gamma}{m}\,\widetilde L^{(G)}
\end{bmatrix}.
\end{equation}

{The following theorem justifies the use of \(\hat{\mathcal C}_\gamma\) for approximating the null distribution.

\begin{theorem}[Null distribution and plug-in calibration with estimated \(p_1\)]
\label{thm:null-estimated-f1}
Under Assumptions~\ref{ass:kernel}--\ref{ass:bochner}, if \(\gamma_{n,m}\to\gamma\in(0,1)\), then under the null hypothesis, the test statistic satisfies
\begin{equation}
    S_{n,m} \cid 2\|Z_\gamma\|_{\mathcal H}
\;\stackrel{d}{=}\;
2\sum_{k\ge1}\lambda_k\chi^2_{1,k},
\end{equation}
where \(Z_\gamma\sim N_{\mathcal H}(0,C_\gamma)\) and \((\lambda_k)_{k\ge1}\) are the eigenvalues of \(C_\gamma\).
Also, assume (i) there is a compact interval \(\mathcal I\subset \mathbb R\) such that
$
\mathbb P\!\left(\hat\theta(Y_{0i}^\ast)\in \mathcal I \text{ for all } 1\le i\le n\right)\to 1,
$ and (ii) \(p_1\) is bounded away from zero on \(\mathcal I\) 
and a density estimator \(\hat p_1\) of $p_1$ converges to \(p_1\) uniformly on \(\mathcal I\), that is,
$
\sup_{u\in\mathcal I}|\hat p_1(u)-p_1(u)| \cip 0,
\ \inf_{u\in\mathcal I} p_1(u)>0.
$ Then, $$\|\hat C_\gamma-C_\gamma\|_1 \cip 0, \quad \text{ and } \quad \sum_{k\ge1}|\hat\lambda_k-\lambda_k| \cip 0,$$ where \((\hat\lambda_k)_{k\ge1}\) are the eigenvalues of \(\hat C_\gamma\). Hence, if \(r_{n,m}\to\infty\) and \(\sum_{k>r_{n,m}}\hat\lambda_k=o_{\mathbb P}(1)\), then $$\hat c_{1-\alpha} \cip c_{1-\alpha},$$ where \(\hat c_{1-\alpha}\) is the \((1-\alpha)\)-quantile of 
$
2\sum_{k=1}^{r_{n,m}}\hat\lambda_k\chi^2_{1,k},
$ and $\mathbb P_{H_0}(S_{n,m}>\hat c_{1-\alpha})\to\alpha.$
\end{theorem}

\paragraph{Estimated critical value and p-value.}

Given the estimated eigenvalues $\hat\lambda_1,\ldots,\hat\lambda_r$, define
$\hat c_{1-\alpha}$ as the $(1-\alpha)$ quantile of
$
\hat Q:=2\sum_{k=1}^r \hat\lambda_k\,\chi^2_{1,k}.
$
The quantile $\hat c_{1-\alpha}$ is computed using the Imhof formula
\eqref{eq:Imhof} with $(\lambda_k)_{k\le r}=(\hat\lambda_k)_{k\le r}$.
We reject $H_0$ if $\mathcal S_{n,m}>\hat c_{1-\alpha}$.
Under $H_0$ and mild regularity conditions (e.g., uniform consistency of
$\hat\theta$ and $\hat p_1$ on the relevant compact sets),
$\hat{\mathcal C}_\gamma\to\mathcal C_\gamma$ in trace norm, implying
$\hat c_{1-\alpha}\to c_{1-\alpha}$ and asymptotic size $\alpha$. For observed \(S_{n,m}=s\), define
$
\hat p
=
\mathbb P\!\left(2\sum_{k=1}^r \hat\lambda_k\,\chi^2_{1,k}\ge s\right),
$
which can be evaluated via the Imhof formula. Under \(H_0\), \(\hat p \cid \mathrm{Unif}(0,1)\).} The numerical procedure is summarized in Algorithm \ref{algo1}. \\

{
\begin{algorithm}[H]
\label{algo1}
\DontPrintSemicolon
\SetAlgoLined
\SetNlSty{textbf}{}{}
\SetKwInOut{KwIn}{Input}
\SetKwInOut{KwOut}{Output}

\caption{Data–driven RKHS test for distributional equality}
\label{alg:dist-null}

\KwIn{
Control samples $\{(Y_0^j,Y_1^j)\}_{j=1}^m$;
treatment samples $\{(Y_0^{\ast i},Y_1^{\ast i})\}_{i=1}^n$;
kernel $\kappa$ (with derivatives); and
significance level $\alpha$.
}
\KwOut{
Test statistic $\mathcal S_{n,m}$;
critical value $\hat c_{1-\alpha}$; and
$p$–value $\hat p$.
}

\textbf{1. Estimate transport map (control group).}\;
Compute $\hat F_{\mu_0}$ and $\hat F_{\mu_1}$ from
$\{Y_0^j,Y_1^j\}_{j=1}^m$, and set
$\hat\theta := \hat F_{\mu_1}^{-1}\!\circ \hat F_{\mu_0}$.\;

\textbf{2. Construct RKHS features and weights.}\;
For $i=1,\dots,n$, define
$\hat\xi_i := \kappa(\cdot,\hat\theta(Y_0^{\ast i}))
-\kappa(\cdot,Y_1^{\ast i})$.\;
For $j=1,\dots,m$, compute $\hat w_j(Y_0^{\ast i})$ and define
$
\hat\zeta_j
:=
\frac{1}{n}\sum_{i=1}^n
\hat w_j(Y_0^{\ast i})\,
\partial_2\kappa(\cdot,\hat\theta(Y_0^{\ast i}))$.

\textbf{3. Form centered Gram matrices.}\;
Construct and center $\hat K^{(F)}$, $\hat L^{(G)}$, $\hat C^{(FG)}$ as in
Section~\ref{sec:est:null} 
to obtain
$\widetilde K^{(F)}$, $\widetilde L^{(G)}$, $\widetilde C^{(FG)}$.\;

\textbf{4. Assemble the combined operator and spectrum.}\;
Form the Gram matrix $\hat G$ as in \eqref{eq:Gram} and compute its
eigenvalues $\hat\lambda_1\ge\hat\lambda_2\ge\cdots\ge 0$.\;

\textbf{5. Compute test statistic.}\;
Compute 
$\hat\mu := n^{-1}\sum_{i=1}^n \hat\xi_i$ and set $\mathcal S_{n,m}
=
S_{n,m}
:=
2\rho_{n,m}\,\|\hat\mu\|_{\mathcal H}^{2}$ with $\rho_{n,m}:=\frac{nm}{n+m}.$ Equivalently,
$\|\hat\mu\|_{\mathcal H}^{2}
=\frac{1}{n^{2}}\sum_{i,\ell=1}^n
\langle\hat\xi_i,\hat\xi_\ell\rangle_{\mathcal H}
=\frac{1}{n^{2}}\mathbf 1^\top \hat K^{(F)}\mathbf 1$.\;

\textbf{6. Calibrate and report inference.}\;
Using $(\hat\lambda_k)_{k\le r}$, compute $\hat c_{1-\alpha}$ via~\eqref{eq:Imhof}
and the corresponding $p$–value $\hat p$.\;
\end{algorithm}
}

\subsection{Asymptotic distribution under contiguous local alternatives}
\label{sec:dist:local_alt}

We next derive the weak limit of the proposed statistic under a sequence of contiguous alternatives on the treated group. Unlike under the null hypothesis, the first-order term of the statistical functional $T(F \lo n, G \lo m)$   does not vanish under the local alternative sequence. The resulting von-Mises expansion therefore produces a noncentral Gaussian quadratic form.

{To study local power and consistency, we employ a sequence of alternatives approaching $H_0$ through shifts of the structural map. Let $\theta_0=\theta(G_0)$ be fixed and $c_n \to 0$. We consider
\begin{equation}
\label{eq:local_alt_unified}
\mathsf H_{1}^{(n)}:\qquad 
\mu_1^\ast = (\theta_0) \lo {\#} \mu_0^\ast + c_n,
\end{equation}
and two asymptotic regimes for the local alternatives: (i) \emph{Pitman (contiguous) regime} that $c_n = C/\sqrt{n}$ with $C\neq 0$, yielding nontrivial local power and a noncentral Gaussian limit; and (ii) \emph{Moderate-deviation regime} that $\rho_{n,m}c_n^2 \to \infty$, yielding consistency of the test.

Under \eqref{eq:local_alt_unified}, the treated-group law varies with $n$. Define the measurable map $T_n:\mathbb R \to \mathbb R^2$ with $T_n(x):=\bigl(x,\theta_0(x)+c_n\bigr)$. Let $(Y \lo 0 \hi *, Y \lo 1 \hi {*(n)}) = T \lo n (Y \lo 0 \hi *)= ( Y \lo 0 \hi *, \theta \lo 0 ( Y \lo 0 \hi *) + c \lo n )$, where $Y \lo 0 \hi * \sim \mu \lo 0 \hi *$, and let $F \lo 0 \hi {(n)}$ represent the c.d.f. of the random vector $(Y \lo 0 \hi *, Y \lo 1 \hi {*(n)})$. Since the control-group law remains fixed at $G_0$, the relevant expansion is centered at $(F_0^{(n)},G_0)$ rather than at $(F_0,G_0)$.
Define $\mu_D
:=
\mathbb E\!\left[\partial_2\kappa\bigl(\cdot,\theta_0(Y_0^\ast)\bigr)\right]
\in\mathcal H,$ and 
$
J_n
:=
\mathbb E_{F_0^{(n)}}\!\left[
\kappa\bigl(\cdot,\theta_0(Y_0^\ast)\bigr)
-
\kappa(\cdot,Y_1^\ast)
\right]
=
\mathbb E\!\left[
\kappa\bigl(\cdot,\theta_0(Y_0^\ast)\bigr)
-
\kappa\bigl(\cdot,\theta_0(Y_0^\ast)+c_n\bigr)
\right] \in \ca H.$
Here, $J_n$ is the deterministic drift induced by the local alternative. A second-order Taylor expansion of $J_n$ in $\mathcal H$ yields the first-order drift. 

The analysis under \eqref{eq:local_alt_unified} proceeds by separating the effect of the local alternative into a deterministic drift and a stochastic fluctuation. The drift arises from the fact that, under $\mathsf H_{1}^{(n)}$, the population discrepancy between the observed and counterfactual distributions is no longer zero. This population-level deviation is captured by $J_n$, whose leading-term behavior determines the magnitude and direction of the departure from the null.

\begin{lemma}[Drift expansion under local alternatives]
\label{lem:drift_consistent}
Suppose $\sup_y \|\partial_2\kappa(\cdot,y)\|_{\mathcal H}<\infty$ and $\sup_y \|\partial_{22}\kappa(\cdot,y)\|_{\mathcal H}<\infty$. Then
$
J_n=-\,c_n\,\mu_D+O(c_n^2).$
In particular,
$
\|J_n\|_{\mathcal H}^2
=
c_n^2\|\mu_D\|_{\mathcal H}^2+O(c_n^3).
$
\end{lemma}

Lemma~\ref{lem:drift_consistent} characterizes the deterministic drift induced by the local alternative. Our analysis combines this drift with the stochastic fluctuation of the empirical discrepancy. The statistic can be viewed as a second-order functional of the empirical measures, whose leading behavior is obtained by centering at the triangular-array baseline $(F_0^{(n)},G_0)$ and decomposing it into a deterministic term of order $c_n$ and a centered empirical term of order $\rho_{n,m}^{-1/2}$. 

\begin{lemma}[Hilbert-space CLT for the linearized empirical discrepancy]
\label{lem:CLT_local} 
Suppose  Assumptions~\ref{eq:parallel trends}--\ref{ass:bochner} are satisfied.  
For each treatment-group observation $(Y_{0j} \hi *,Y_{1j} \hi * )$, $i=1, \ldots, n$,  define $\Xi_{n,i}
:= 
\kappa(\cdot,Y_{1i}^\ast)-\kappa\bigl(\cdot,\theta_0(Y_{0i}^\ast)\bigr)$ and $J_n:=  \mathbb E(\Xi_{n,1})
=
\mathbb E_{F_0^{(n)}}\!\left[
\kappa(\cdot,Y_1^\ast)-\kappa\bigl(\cdot,\theta_0(Y_0^\ast)\bigr)
\right].$
For each controlled-group observation \((Y_{0j},Y_{1j})\), $j=1, \ldots, m$,  define
$\Psi_j
:= 
\mathbb E\!\left[
w_j(Y_0^\ast)\,\partial_2\kappa\bigl(\cdot,\theta_0(Y_0^\ast)\bigr)
\right]\in\mathcal H$ with $w_j(y)
:= 
1\{Y_{0j}\le y\}-F_{\mu_0}(y)
-
\frac{1\{Y_{1j}\le \theta_0(y)\}-F_{\mu_1}(\theta_0(y))}{p_1(\theta_0(y))}$, where the expectation in the definition of $\Psi \lo j$ is taken with respect to  \(\mu_0^\ast\). Let 
 $\mathbb G_{n,m}
:=  
\sqrt{\rho_{n,m}}
\left[
\frac1n\sum_{i=1}^n\{\Xi_{n,i}-J_n\}
+
\frac1m\sum_{j=1}^m \Psi_j
\right]$. 
If \(n/(n+m)\to\gamma\in(0,1)\), then,  under the local alternative \eqref{eq:local_alt_unified}, 
\(
\mathbb G_{n,m}\rightsquigarrow Z_\gamma \in \mathcal H,
\)
where $\rightsquigarrow$ represents weak convergence in Hilbert space,  \(Z_\gamma\) is the same mean-zero Gaussian random element as in Theorem~\ref{thm:main}, with covariance operator
$
\mathcal C_\gamma=(1-\gamma)\mathcal C^{(F)}+\gamma \mathcal C^{(G)}.
$
Equivalently, for every \(h\in\mathcal H\), we have
$
\big\langle \mathbb G_{n,m},h\big\rangle_{\mathcal H}
\; \cid \;
N\!\left(0,\,
\big\langle \mathcal C_\gamma h,h\big\rangle_{\mathcal H}\right).
$
\end{lemma}

Lemma \ref{lem:CLT_local} identifies the limiting distribution of this centered $\mathcal H$-valued empirical component. Taken together, Lemmas~\ref{lem:drift_consistent} and \ref{lem:CLT_local} show that, after centering at $(F_0^{(n)},G_0)$, the empirical discrepancy consists of a deterministic component of order $c_n$ and a centered Gaussian fluctuation of order $\rho_{n,m}^{-1/2}$. This decomposition directly yields the distinct limiting regimes described in the following theorems.
We now state the contiguous-limit result.

\begin{theorem}[Contiguous local alternatives]
\label{thm:local-power}
Suppose that Assumptions~\ref{eq:parallel trends}--\ref{ass:bochner} are satisfied. Let
$
\Delta_\gamma:=-\sqrt{1-\gamma}\,C\,\mu_D.
$ If $n/(n+m)\to\gamma\in(0,1)$ and $c \lo n = C / \sqrt n$ for some $C \ne 0$, then
\begin{equation}
\label{eq:pitman-limit}
S_{n,m} \cid 2\|Z_\gamma+\Delta_\gamma\|_{\mathcal H}^2,
\end{equation}
where $Z_\gamma\sim N(0,\mathcal C_\gamma)$ is the mean-zero Gaussian element in $\mathcal H$ with covariance operator $\mathcal C_\gamma=(1-\gamma)\mathcal C^{(F)}+\gamma\mathcal C^{(G)}$ as in Theorem~\ref{thm:main}. Hence, if $\hat c_{1-\alpha}\to c_{1-\alpha}$ in probability, where $c_{1-\alpha}$ denotes the $(1-\alpha)$-quantile of $2\|Z_\gamma\|_{\mathcal H}^2$, then we have
\[
\lim\limits_{n,m\to\infty}
\mathbb P_{\mathsf H_{1}^{(n)}}\!\bigl(S_{n,m}>\hat c_{1-\alpha}\bigr)
=
\mathbb P\!\bigl(2\|Z_\gamma+\Delta_\gamma\|_{\mathcal H}^2>c_{1-\alpha}\bigr).
\]
\end{theorem}

In the Pitman regime, the deterministic drift induced by the alternative is of the same order as the stochastic fluctuations of the empirical process. Consequently, the statistic converges to a noncentral quadratic form in a Gaussian element of $\mathcal H$, yielding a nontrivial limiting power strictly between $\alpha$ and $1$. The noncentrality parameter is governed by 
the directional derivative of the kernel mean map along the structural shift, and its projection onto the eigenspaces of the covariance operator $\mathcal C_\gamma$. Thus, local power depends on how well the alternative aligns with directions of high variance in the RKHS, reflecting the interaction between the kernel and the geometry of the alternative, and the asymptotic rejection probability is, in general, strictly between $\alpha$ and $1$.  Moreover, we develop a feasible plug-in approximation procedure for the asymptotic local alternative distribution given in Theorem \ref{thm:local-power} in the supplement.

We next study alternatives that diverge from $H_0$ more slowly than the contiguity rate.

\begin{theorem}[Consistency beyond the contiguous regime]
\label{thm:local-consistency}
Under Assumptions~\ref{eq:parallel trends}--\ref{ass:bochner}, if $\rho_{n,m}c_n^2\to\infty$,
and $\|\mu_D\|_{\mathcal H}>0$, then
$
S_{n,m} \cip \infty.
$
Consequently, we have
$$
\mathbb P_{\mathsf H_{1}^{(n)}}\!\bigl(S_{n,m}>\hat c_{1-\alpha}\bigr)\to 1.
$$
\end{theorem}
Theorem~\ref{thm:local-consistency} shows that, under moderate deviations that $\rho_{n,m}c_n^2\to\infty$, the deterministic drift dominates the $O_{\mathbb P}(1)$ stochastic terms in the second-order expansion. The statistic diverges in probability, and the rejection probability converges to one. Hence, the test is consistent against any sequence of alternatives that deviates from the null at a rate exceeding the $1/\sqrt n$ scale. Together, the two regimes delineate the precise detection boundary of the procedure: $c_n\asymp n^{-1/2}$ yields nontrivial local power, while any faster rate ensures consistency.}

\subsection{Finite-sample guarantee}
\label{sec:finite-sample}

 This section provides an explicit nonasymptotic bound for the proposed V-statistic $V \lo {n,m} = T ( F \lo n, G \lo m)$ in~\eqref{eq:V-stat}. The bound controls
$
\bigl|\,T(F_n,G_m)-T(F_0,G_0)\,\bigr|
$
and decomposes into two error sources:
(i) treated-sample fluctuation around $T(F_0,G_m)$, controlled by bounded-kernel concentration for the 
V-statistics, and (ii) the plug-in error $T(F_0,G_m)-T(F_0,G_0)$ induced by estimating the monotone
transport map \(\theta(G_0)\) by the empirical drift map \(\theta(G_m)\), estimated from the control group samples, controlled via DKW-type uniform laws~\citep{dvoretzky1956asymptotic}, and the
stability of quantile maps under a density lower bound. 

\begin{assumptions}[Bounded and coordinate-wise Lipschitz kernel]
\label{ass:kernel2_main}
There exist constants \(K\in(0,\infty)\) and \(L\in(0,\infty)\) such that
$\sup_{a,b\in\mathbb R}|k(a,b)|\le K$, 
$|k(a,b)-k(a',b)|\le L|a-a'|$, and
$|k(b,a)-k(b,a')|\le L|a-a'|$ 
for all \(a,a',b\in\mathbb R\).
\end{assumptions}

\begin{assumptions}[Density lower bound for the control post-period marginal]
\label{ass:density_main}
The control post-period marginal \(\mu_1\) has density \(p_1\) satisfying
$0<p_{1,\min}\le p_1(y)$
for all $y\in\mathbb R$.
\end{assumptions}

We will work under Assumptions \ref{ass:kernel2_main}--\ref{ass:density_main} that are standard in the literature. 

\begin{theorem}[Finite-sample bound for the plug-in V-statistic]
\label{thm:finite_sample_V_supp}
Suppose that Assumptions~\ref{ass:kernel2_main} and \ref{ass:density_main} are satisfied. Then for any \(t>0\) and \(\varepsilon>0\), we have
\begin{align}
\label{eq:finite_sample_tail_ms_V}
\begin{split}
\ali \PP\!\left(
\big|V \lo {n,m}^2 - T(F_0,G_0)\big|
>
t + \frac{8K}{n} + \big(8L\sqrt K + 4L^2\varepsilon\big)\varepsilon
\right) \\
\ali \hspace{1in} \le
2\exp\!\left(-\frac{n t^2}{128K^2}\right)
+
4\exp\!\left(-2m p_{1,\min}^2(\varepsilon/2)^2\right).
\end{split}
\end{align}
Equivalently, let $\varepsilon_\delta:=\frac{1}{p_{1,\min}}\sqrt{\frac{2}{m}\log\frac{8}{\delta}}$, and for any \(\delta\in(0,1)\), with probability at least \(1-\delta\), 
\begin{align}
\label{eq:finite_sample_hp_ms_V}
\big|V\lo {n,m}^2 - T(F_0,G_0)\big|
\le
\sqrt{\frac{128K^2}{n}\log\frac{4}{\delta}}
+
\frac{8K}{n}
+
\Big(8L\sqrt K+4L^2\varepsilon_\delta\Big)\varepsilon_\delta.
\end{align}
\end{theorem}

Theorem~\ref{thm:finite_sample_V_supp} separates the treated-sample fluctuation and drift-map estimation error. The first term in \eqref{eq:finite_sample_tail_ms_V} decays at the rate \(\exp(-c n t^2)\), reflecting the concentration of the V-statistic around its conditional mean, while the second term decays at the rate \(\exp(-c' m \varepsilon^2)\), arising from estimation of \(\theta(G_m)\). The factor
$
\bigl(8L\sqrt K+4L^2\varepsilon\bigr)\varepsilon
$
captures how the first- and second-order drift-map estimation errors propagate through the squared RKHS norm.

\section{Simulation Studies}
\label{sec:sim}
This section evaluates the finite-sample performance of the proposed distribution-based test and compares it with the classical difference-in-differences (DiD) procedure. We consider three designs that stress different aspects of distributional inference: heavy tails, mixture-induced heterogeneity, and shape-only deviations. Across all experiments, the nominal level is \(\alpha=0.05\), and rejection frequencies are computed from \(200\) Monte Carlo replications.

In the one-dimensional settings considered here, the population transport map is
\(
\theta_0(x)=F_{\mu_1}^{-1}\!\bigl(F_{\mu_0}(x)\bigr),
\)
that is, \(\theta_0=\theta(G_0)\) in the notation of Section~\ref{sec:problem}. In each replication, an independent control sample of size \(m\) is generated from \((\mu_0,\mu_1)\), and the empirical transport map \(\widehat\theta_m=\theta(G_m)\) is constructed by empirical optimal transport with barycentric projection. The proposed test targets the transport-compatibility null
\(
H_0:\ \mu_1^\ast=\theta_0\#\mu_0^\ast,
\)
where \(\mu_0^\ast\) and \(\mu_1^\ast\) are the pre- and post-treatment marginals in the treated group. 

To implement this null, conditional on \(\widehat\theta_m\), we generate two independent samples
$\{Y_{0i}^\ast\}_{i=1}^n$, $\{Z_i\}_{i=1}^n$
from \(\mu_0\). The counterfactual sample is \(\{\widehat\theta_m(Y_{0i}^\ast)\}_{i=1}^n\). Under \(H_0\), the post-treatment sample is generated as
$Y_{1i}^\ast=\widehat\theta_m(Z_i)$, $i=1,\dots,n$,
so that \(Y_1^\ast\) and \(\widehat\theta_m(Y_0^\ast)\) have the same conditional law. Thus the null is imposed at the distributional level, without requiring a pathwise identity between \(Y_{0i}^\ast\) and \(Y_{1i}^\ast\). Power is studied under the local-alternative framework of Section~\ref{sec:dist:local_alt}. There, departures of magnitude \(c_n\) yield, at the Pitman boundary \(c_n=C/\sqrt n\), the noncentral limiting shift
$\Delta_\gamma=-\sqrt{1-\gamma}\,C\,\mu_D$. 
In the simulations, this local-deviation mechanism is implemented by
$Y_{1i}^\ast
=
\widehat\theta_m(Z_i)+\delta_n\,\sign(Z_i)$ with
$\delta_n=Cn^{-r}$ and $0<r<\tfrac12$. 
Hence, \(\delta_n\) is the finite-sample analogue of the deviation magnitude \(c_n\). Choosing \(r<1/2\) gives a vanishing but stronger-than-contiguous perturbation, corresponding to the moderate-deviation regime in which the proposed test is expected to be consistent.

\textbf{DiD benchmark.}
For the DiD benchmark, we generate a separate two-period sample from the same baseline law. Let \(U_{0i},U_{1i}\stackrel{\mathrm{i.i.d.}}{\sim}\mu_0\), and define
$Y_{0i}=\widehat\theta_m(U_{0i})+\varepsilon_{0i}$, and
$Y_{1i}=\widehat\theta_m(U_{1i})+\varepsilon_{1i}
+\delta_n\,\sign(U_{1i})$, where \(\varepsilon_{0i}\) and \(\varepsilon_{1i}\) are independent Gaussian noise terms with scenario-specific variance. Independent labels \(G_i\sim\mathrm{Bernoulli}(1/2)\) are assigned, and DiD is computed as the difference in average changes \(Y_{1i}-Y_{0i}\) between the two groups. Since \(G_i\) is independent of all outcomes, both groups have the same mean change by construction. The DiD estimand is therefore zero under both \(H_0\) and \(H_{1,n}\), so this benchmark has no group-specific mean signal to exploit.

The scenarios below differ only in the specification of \((\mu_0,\mu_1)\), allowing us to examine how the proposed method behaves under increasingly challenging distributional features.

\textbf{Scenario I: Non-normal design.}
We take $\mu_0=t_3$, and $\mu_1=\mathrm{LogNormal}(0,1)$. This yields a heavy-tailed baseline law, a highly skewed post-period law, and a nonlinear transport map. The DiD sample uses Gaussian panel noise with variance \(0.8^2\). 

\textbf{Scenario II: Mixture distributions and heterogeneity.}
Take
$\mu_0
=
0.55\,N(-1,0.7^2)+0.45\,N(1.4,0.9^2)$,
and
$\mu_1
=
0.40\,\mathrm{LogNormal}(0,0.5^2)
+
0.60\,\mathrm{LogNormal}(1.1,0.6^2)$. 
The DiD sample uses Gaussian panel noise with variance \(0.6^2\). This scenario introduces multimodality, skewness, and component-specific scale variation.

\textbf{Scenario III: Shape-only alternatives with mean preservation.}
We take $\mu_0=N(0,1)$ and $\mu_1=N(3,1)$ for which the population transport map is linear. The DiD sample uses Gaussian panel noise with variance \(1\). Under \(H_{1,n}\), the perturbation changes the shape of the post-treatment law while preserving its first moment, isolating the contrast between distribution-based and mean-based inference.

\textbf{Results.}
Table~\ref{tab:sim-size-power} reports empirical size and power. The proposed test has an empirical size under \(H_0\) close to the nominal level for all the settings. Under \(H_{1,n}\), it attains high power and approaches power one as \(n\) increases. By contrast, the DiD rejection frequencies remain close to the nominal level, consistent with the absence of a group-specific mean signal.

\begin{table}[htbp]
\centering
\caption{Empirical size (under \(H_0\)) and power (under \(H_{1,n}\)) of the distribution-based test and the classical difference-in-differences (DiD) test across three simulation scenarios. Rejection frequencies are computed over \(200\) Monte Carlo replications at level \(\alpha=0.05\). 
}
\label{tab:sim-size-power}
\begin{tabular}{llcccc}
\toprule
Scenario & \(n\) 
& \multicolumn{2}{c}{Empirical size (under \(H_0\))} 
& \multicolumn{2}{c}{Empirical power (under \(H_{1,n}\))} \\
\cmidrule(lr){3-4} \cmidrule(lr){5-6}
& & Distributional & DiD & Distributional & DiD \\
\midrule
\multirow{3}{*}{I}
& 50   & 0.0362 & 0.0461 & 0.9975 & 0.0576 \\
& 300  & 0.0473 & 0.0515 & 1.0000 & 0.0489 \\
& 1000 & 0.0323 & 0.0264 & 1.0000 & 0.0510 \\
\midrule
\multirow{3}{*}{II}
& 50   & 0.0500 & 0.0550 & 0.7500 & 0.0495 \\
& 300  & 0.0475 & 0.0550 & 0.8765 & 0.0515 \\
& 1000 & 0.0500 & 0.0625 & 1.0000 & 0.0498 \\
\midrule
\multirow{3}{*}{III}
& 50   & 0.05745 & 0.0550 & 0.9825 & 0.0450 \\
& 300  & 0.0475  & 0.0575 & 1.0000 & 0.0425 \\
& 1000 & 0.0350  & 0.0500 & 1.0000 & 0.0400 \\
\bottomrule
\end{tabular}
\end{table}

\begin{figure}[htbp]
\centering
\caption{Finite-sample calibration across the three simulation scenarios. Columns correspond to Scenarios~I--III. The top row displays the empirical null distribution of the statistic under \(H_0\), with the observed statistic (dotted vertical line) and the critical value (dashed vertical line). The bottom row overlays the statistic from an independent replicate generated under \(H_{1,n}\), showing the displacement of the observed statistic toward the upper tail.}
\label{fig:sim-perm-overlay}
\vspace{0.5em}

\begin{tabular}{ccc}
\includegraphics[width=0.31\textwidth]{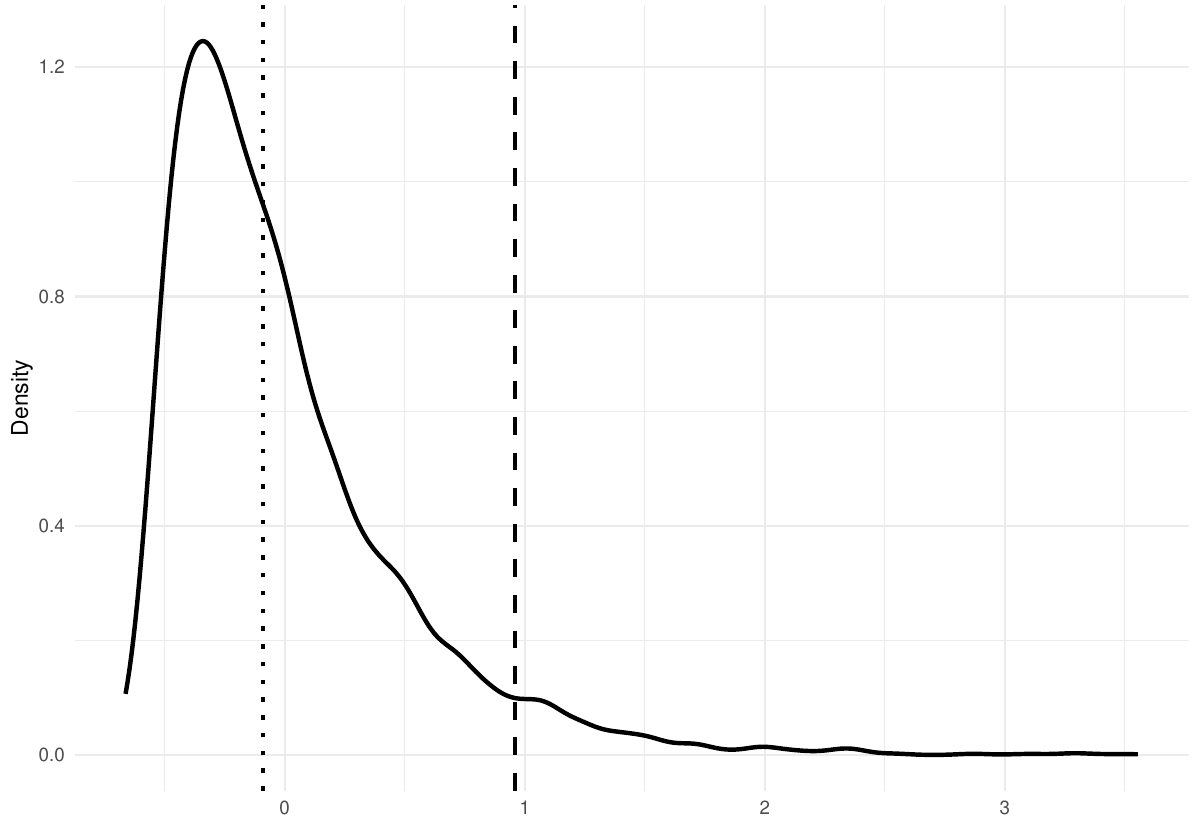} &
\includegraphics[width=0.31\textwidth]{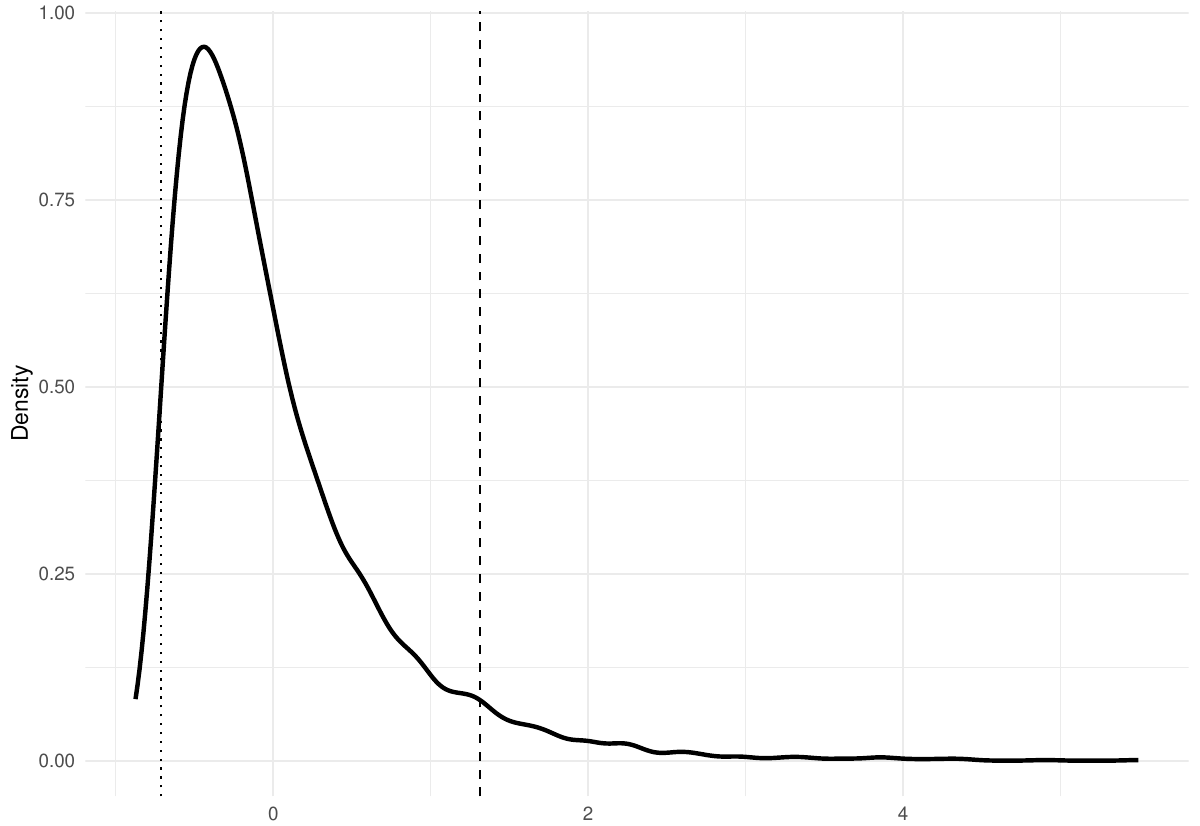} &
\includegraphics[width=0.31\textwidth]{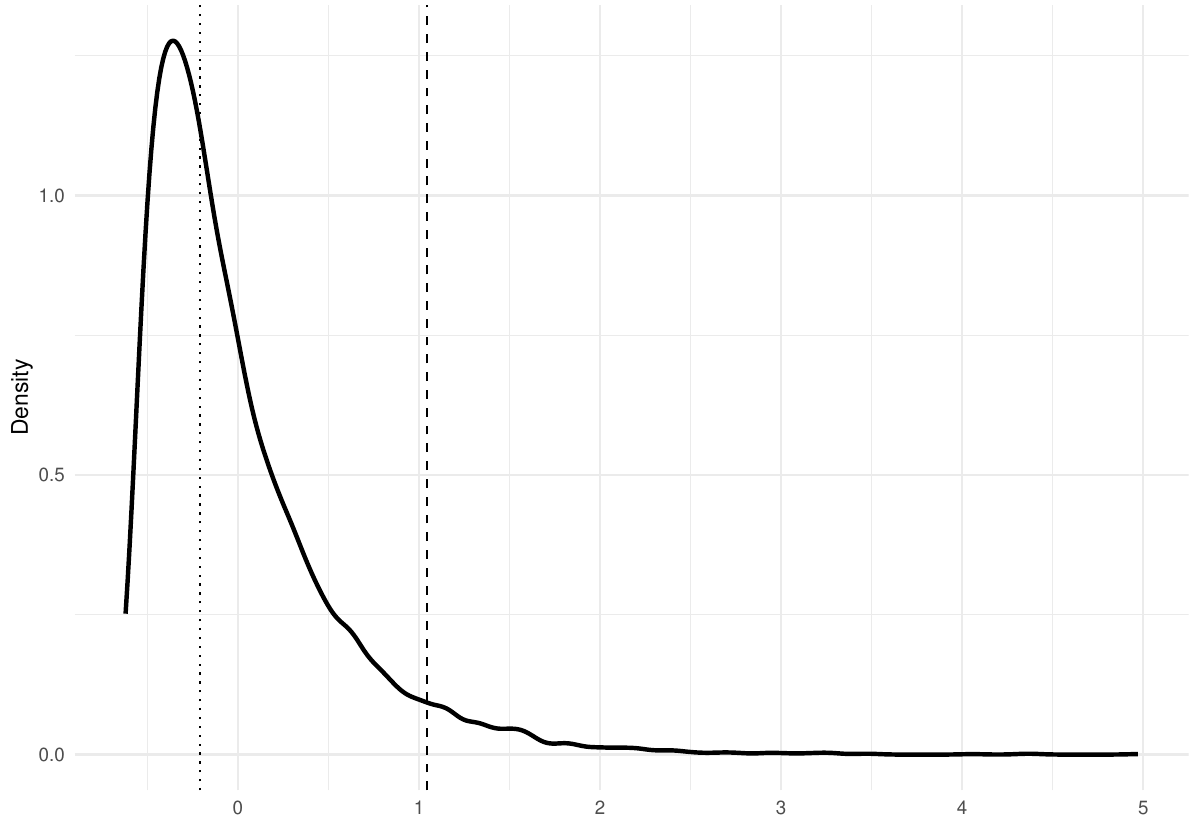} \\
\small Scenario I: \(H_0\) &
\small Scenario II: \(H_0\) &
\small Scenario III: \(H_0\) \\[0.6em]
\includegraphics[width=0.31\textwidth]{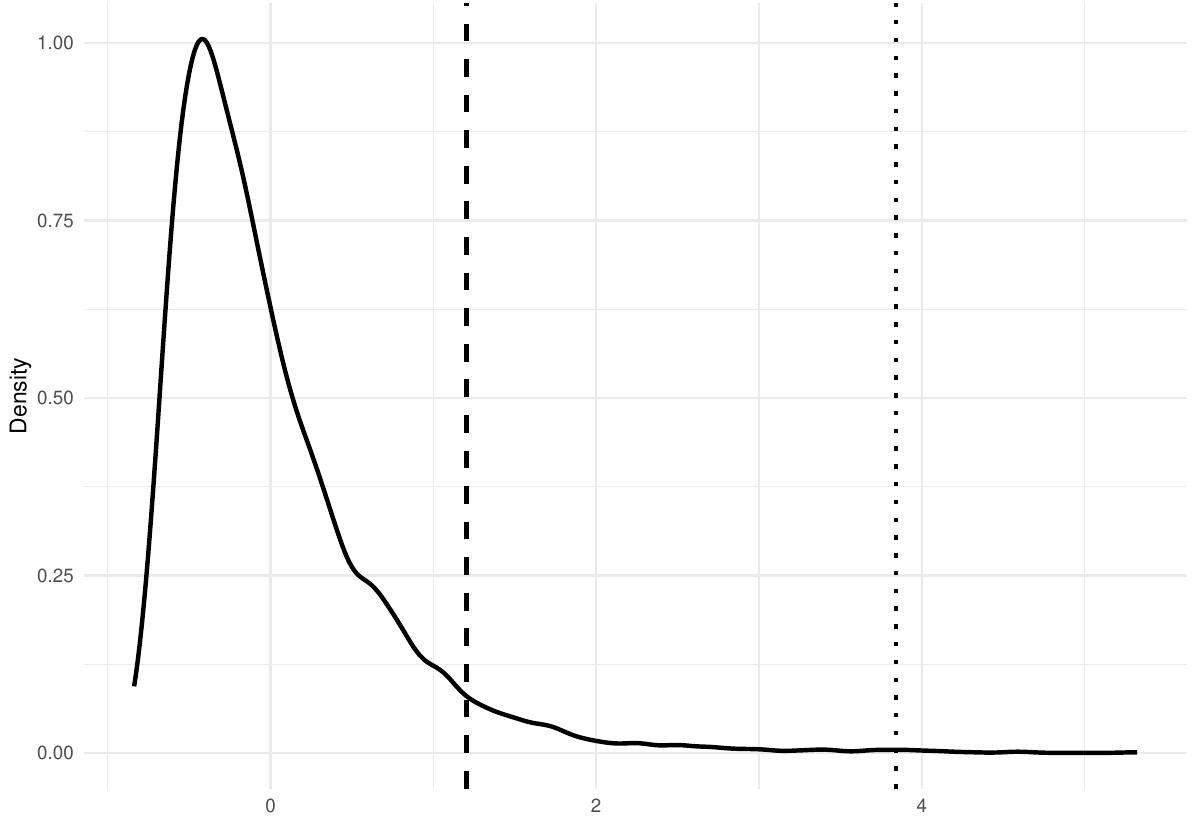} &
\includegraphics[width=0.31\textwidth]{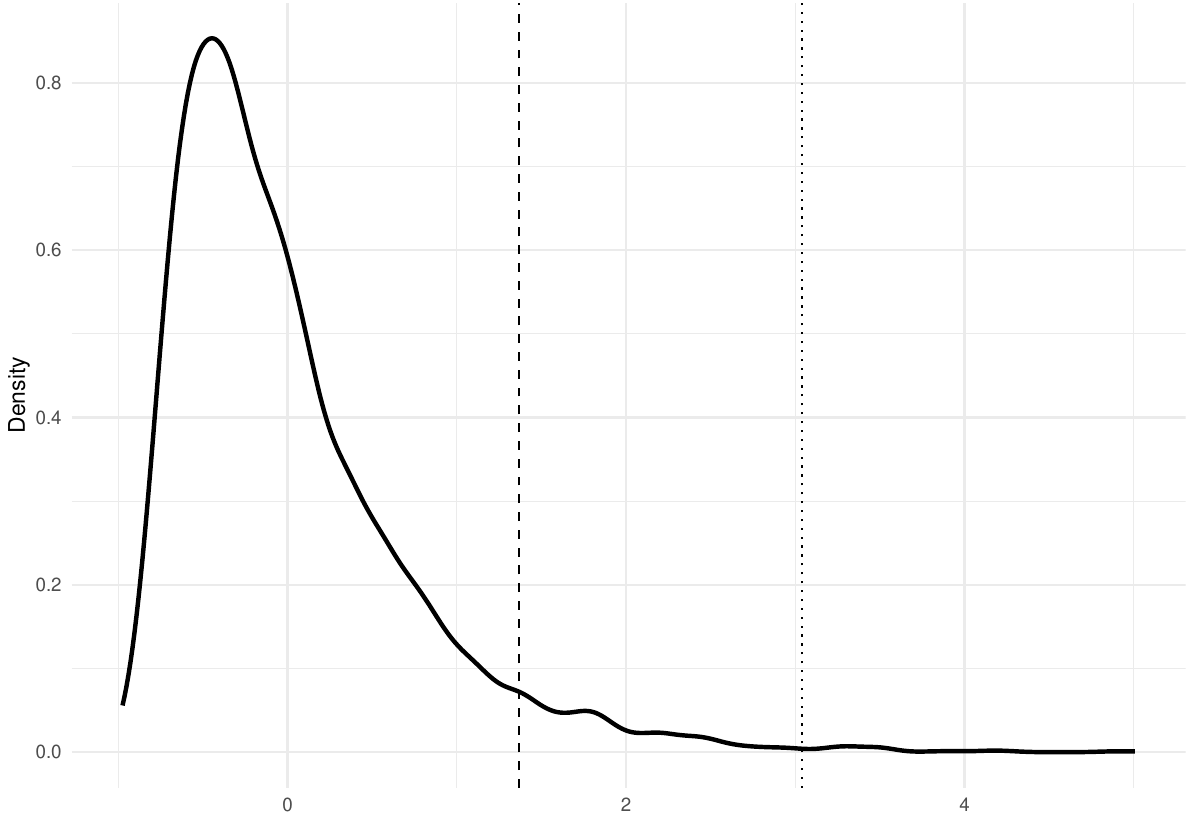} &
\includegraphics[width=0.31\textwidth]{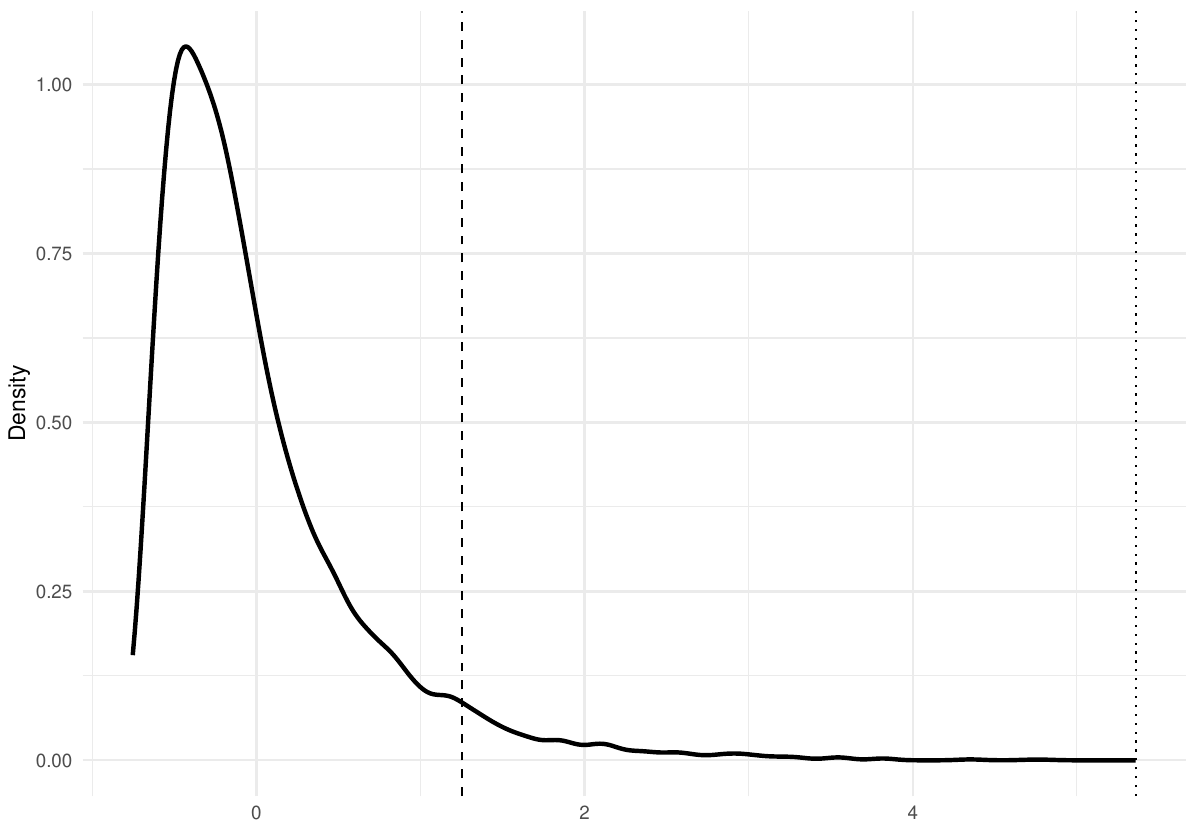} \\
\small Scenario I: \(H_{1,n}\) &
\small Scenario II: \(H_{1,n}\) &
\small Scenario III: \(H_{1,n}\) \\
\end{tabular}
\end{figure}

\begin{figure}[htbp]
\caption{Empirical power as a function of the treatment sample size \(n\) under the local alternative \(H_{1,n}\). In each panel, the solid blue curve corresponds to the distribution-based test and the dashed curve corresponds to the classical DiD test. Power is computed as the Monte Carlo rejection frequency at level \(\alpha=0.05\) over \(200\) replications.}
\label{fig:sim-power-consistency}
\begin{minipage}[t]{0.33\textwidth}
\centering
\includegraphics[width=\textwidth]{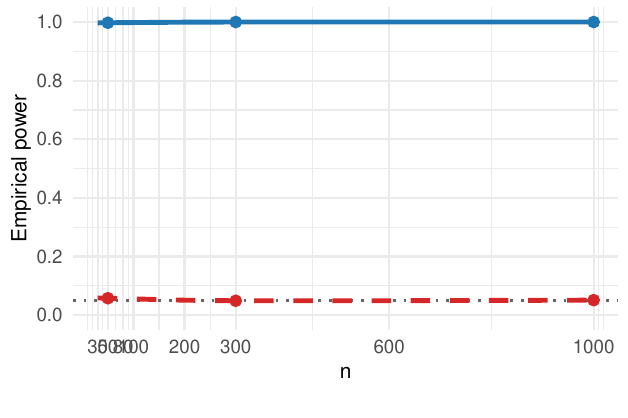}
\vspace{-0.5em}
\par\small (A) Scenario I.
\end{minipage}\hfill
\begin{minipage}[t]{0.33\textwidth}
\centering
\includegraphics[width=\textwidth]{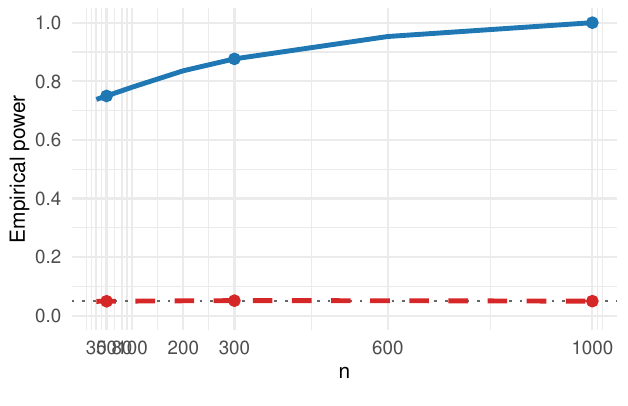}
\vspace{-0.5em}
\par\small (B) Scenario II.
\end{minipage}\hfill
\begin{minipage}[t]{0.33\textwidth}
\centering
\includegraphics[width=\textwidth]{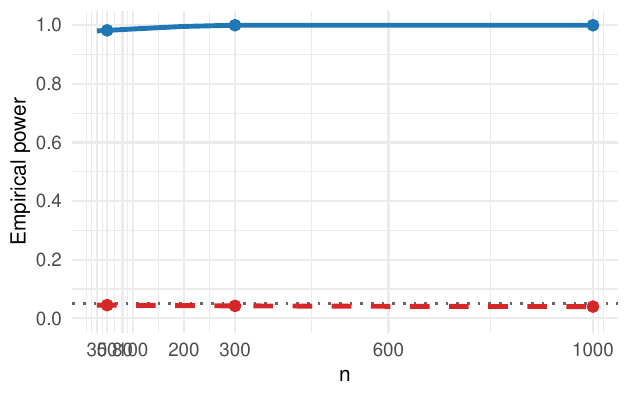}
\vspace{-0.5em}
\par\small (C) Scenario III.
\end{minipage}
\end{figure}

Figure~\ref{fig:sim-perm-overlay} visualizes the finite-sample calibration mechanism. The observed statistic lies within the bulk of the null distribution under \(H_0\) and shifts toward the upper tail and often exceeds the critical threshold under \(H_{1,n}\), illustrating how local distributional departures generate finite-sample separation from the null. Figure~\ref{fig:sim-power-consistency} confirms the power pattern in Table~\ref{tab:sim-size-power}. The proposed test gains power monotonically and approaches one in all scenarios, whereas DiD remains essentially flat near the nominal level. Overall, the simulations show that the proposed test maintains reliable size control and detects local departures from transport compatibility, while classical DiD is insensitive when treatment effects appear through the shape of the outcome distribution rather than a group-specific mean change.

\section{Real Data Analysis}
\label{sec:data}

We demonstrate the utility of the proposed distribution-based test using the seminal minimum-wage study of \citet{card1994minimum}. On April 1, 1992, New Jersey increased its minimum wage from \$4.25 to \$5.05 per hour, while neighboring Pennsylvania maintained the federal minimum wage at \$4.25. This policy change provides a canonical two-group, two-period design for evaluating the employment effects of minimum-wage legislation. 

The dataset consists of 331 fast-food restaurants in New Jersey and 79 in eastern Pennsylvania, which were surveyed in February 1992, before the policy change, and again in November 1992, after the increase had taken effect. We take total restaurant employment as the outcome, with New Jersey serving as the treated group and Pennsylvania as the control group. \citet{card1994minimum} focused on average employment effects and found little evidence that the minimum-wage increase would reduce employment. Our analysis aims to explore whether the policy altered the \emph{distribution} of employment outcomes. Specifically, we test whether the observed post-treatment employment distribution in New Jersey is compatible with the counterfactual distribution obtained by transporting New Jersey's pre-treatment employment distribution according to the distributional drift observed in Pennsylvania.

\begin{figure}[!ht]
    \centering
    \begin{minipage}{0.45\textwidth}
        \centering
        \includegraphics[width=0.8\textwidth]{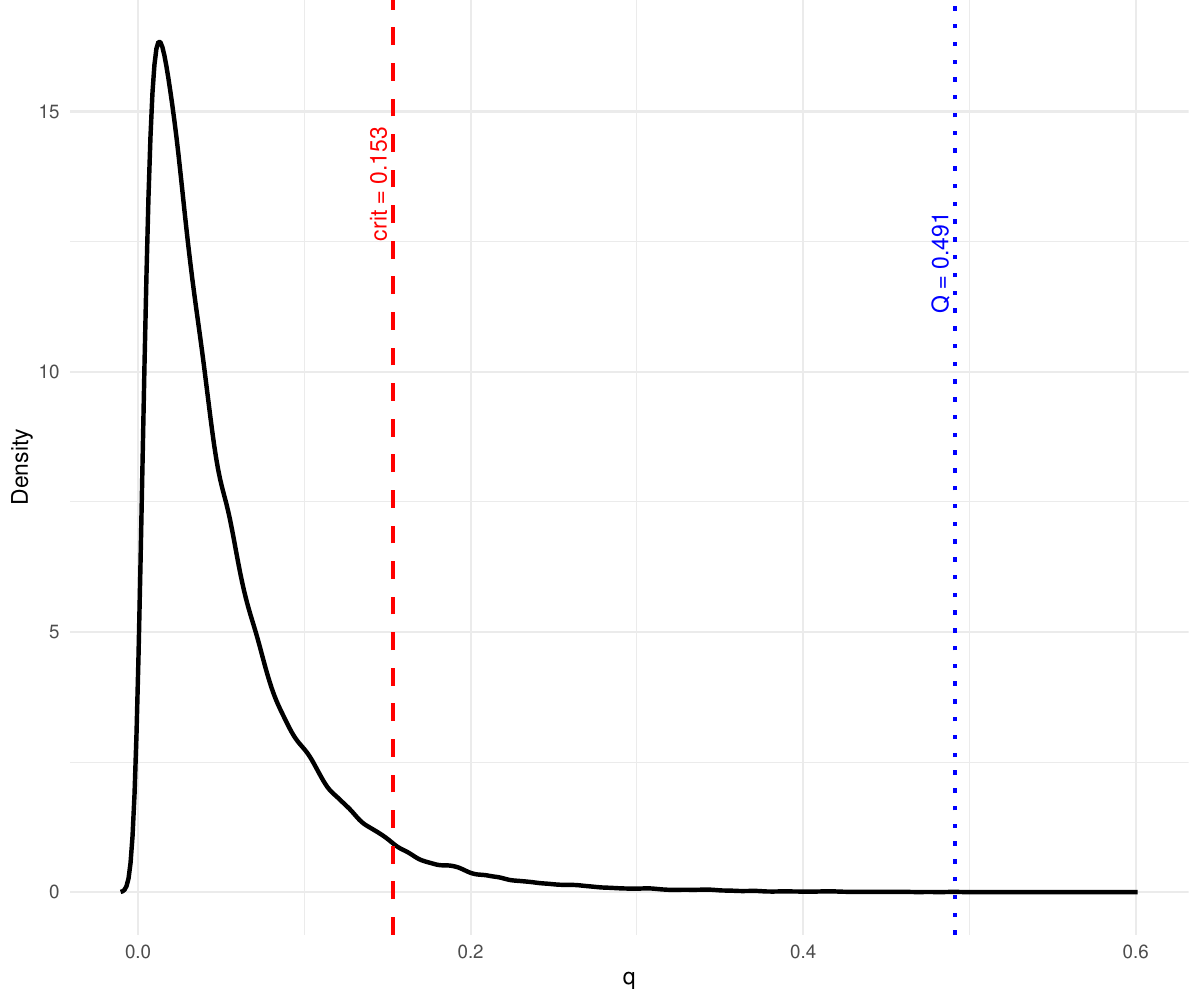}
    \end{minipage}\hfill
    \begin{minipage}{0.45\textwidth}
        \centering
        \includegraphics[width=0.8\textwidth]{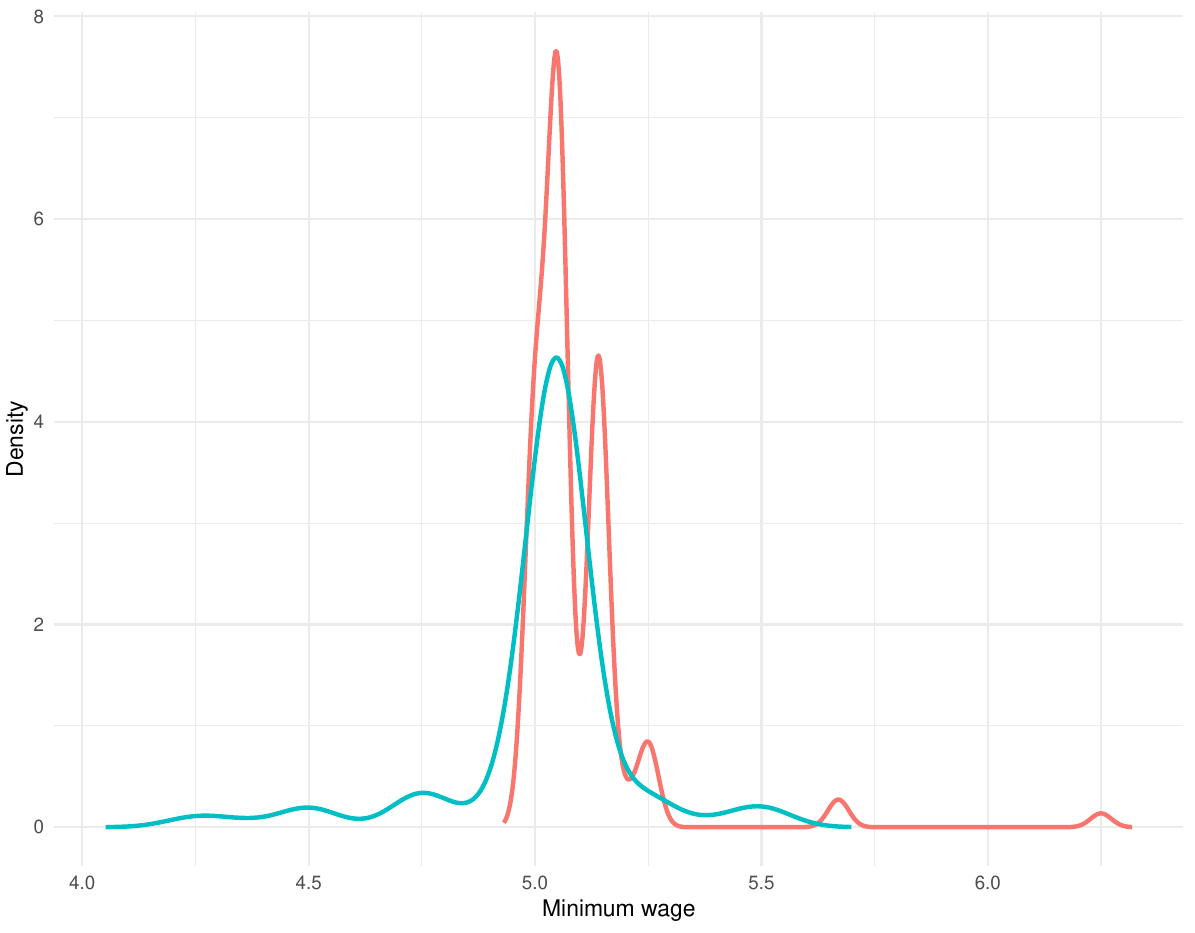}
    \end{minipage}
    \caption{Card--Krueger minimum-wage study. Left: estimated asymptotic null distribution of the distribution-based test statistic, with the observed statistic and critical threshold overlaid. Right: observed post-treatment employment distribution for New Jersey restaurants (in blue) and the counterfactual distribution implied by Pennsylvania controls (in red).}
    \label{fig:card_krueger}
\end{figure}

Figure~\ref{fig:card_krueger} reports the results. In the left panel, the observed test statistic is \(0.4913\), substantially exceeding the critical value \(0.1535\). The corresponding asymptotic \(p\)-value is \(1.7\times 10^{-4}\), with Monte Carlo standard error \(4.1\times 10^{-5}\). We therefore reject the null hypothesis that the post-treatment employment distribution in New Jersey is explained by the control-implied distributional drift. The right panel compares the observed post-treatment distribution in New Jersey with the corresponding counterfactual distribution. Although the two distributions have broadly similar centers, they differ in dispersion and tail behavior. These discrepancies are precisely the type of distributional effects that are not summarized by a mean contrast but are detected by the proposed test. The empirical findings refine the conventional interpretation of the Card--Krueger study. 

\section{Conclusion}
\label{sec:concl}

We introduce a distributional framework for causal inference in two-period designs, extending classical DiD to distribution-based inference by combining counterfactual OT with RKHS-based discrepancy measures. Testing the distributional equality between observed post-treatment and transported counterfactual outcomes yields a nonparametric test sensitive to changes in dispersion, shape, and tail behavior. We provide a complete asymptotic characterization: the test statistic converges to a Gaussian quadratic form under the null, while under local alternatives, we establish a unified theory encompassing Pitman local power and moderate-deviation consistency. This clarifies how detectability depends on the interaction between transport-induced drift and RKHS geometry. Our approach embeds causal structure directly into distributional comparisons while retaining analytical tractability.

{
\bibliographystyle{agsm}
\bibliography{ref}
}

\end{document}